\documentclass[prb,amsmath,superscriptaddress,citeautoscript,twocolumn,showkeys,showpacs,floatfix]{revtex4}
\usepackage{bm}
\usepackage{graphicx}
\usepackage{amssymb}

\newcommand{\spinup}{\uparrow}
\newcommand{\spindown}{\downarrow}
\newcommand{\qav}[1]{\langle {#1} \rangle}

\DeclareMathOperator{\sgn}{sgn}

\DeclareMathOperator{\re}{Re}

\DeclareMathOperator{\sech}{sech}

\begin{document}

\author{Vyacheslavs Kashcheyevs}
\email{slava@latnet.lv}
\affiliation{School of Physics and Astronomy,
             Raymond and Beverly Sackler Faculty of Exact
             Sciences,\\
             Tel Aviv University, Tel Aviv 69978, Israel}

\author{Avraham Schiller}
\affiliation{Racah Institute of Physics,
             The Hebrew University, Jerusalem 91904, Israel}

\author{Amnon Aharony}
\affiliation{Department of Physics, Ben Gurion University,
             Beer Sheva 84105, Israel}

\author{Ora Entin-Wohlman}
\affiliation{Department of Physics, Ben Gurion University,
             Beer Sheva 84105, Israel}

\affiliation{Albert Einstein Minerva Center for Theoretical
             Physics, Weizmann Institute of Science,
             Rehovot 76100, Israel}


\title{Unified description of correlations in double quantum dots}
\begin{abstract}

The two-level model for a double quantum dot coupled to two
leads, which is ubiquitously used to describe charge
oscillations, transmission-phase lapses and correlation-induced
resonances, is considered in its general form. The model
features arbitrary tunnelling matrix elements among the two
levels and the leads and between the levels themselves
(including the effect of Aharonov-Bohm fluxes), as well as
inter-level repulsive interactions. We show that this model is
exactly mapped onto a generalized Anderson model of a single
impurity, where the electrons acquire a pseudo-spin degree of
freedom, which is conserved by the tunnelling but not within
the dot. Focusing on the local-moment regime where the dot is
singly occupied, we show that the effective low-energy
Hamiltonian is that of the anisotropic Kondo model in the
presence of a tilted magnetic field. For moderate values of the
(renormalized) field, the Bethe \emph{ansatz} solution of the
isotropic Kondo model allows us to derive accurate expressions
for the dot occupation numbers, and henceforth its
zero-temperature transmission. Our results are in excellent
agreement with those obtained from the Bethe \emph{ansatz} for
the isotropic Anderson model, and with the functional and
numerical renormalization-group calculations of Meden and
Marquardt [Phys. Rev. Lett. \textbf{96}, 146801 (2006)], which
are valid for the general anisotropic case. In addition we
present highly accurate estimates for the validity of the
Schrieffer-Wolff transformation (which maps the Anderson
Hamiltonian onto the low-energy Kondo model) at both the high-
and low-magnetic field limits. Perhaps most importantly, we
provide a single coherent picture for the host of phenomena to
which this model has been applied.
\end{abstract}



\maketitle

\section{Introduction}

The ongoing technological progress in the fabrication and
control of nanoscale electronic circuits, such as quantum dots,
has stimulated detailed studies of various quantum-impurity
models, where a few local degrees of freedom are coupled to a
continuum. Of particular interest are models with
experimentally verifiable universal properties. One of the best
studied examples is the Anderson single impurity
model,~\cite{Anderson61} which describes successfully
electronic correlations in small quantum
dots~\cite{NgLee88,GlazmanRaikh88}. The experimental control of
most of the parameters of this model, e.g., the impurity energy
level position or the level broadening due to hybridization
with the continuum, allows for detailed
investigations~\cite{DGG98,vanderWiel00} of the universal
low-temperature behavior of the Anderson model.

\begin{figure*}
\includegraphics[width=14cm]{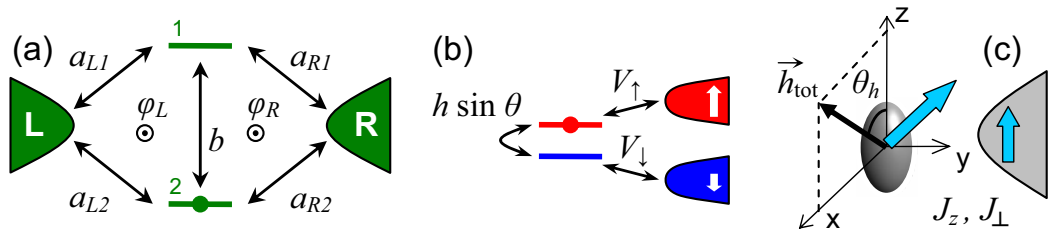}
\caption{A schematic representation of the double-dot system,
         along with its reduction in the local-moment regime
         to an effective Kondo model with a tilted magnetic
         field.
         (a) The model system: two localized levels coupled
         by tunnelling matrix elements to one another and
         to two separate leads. A constant magnetic flux
         induces phase factors on those elements. Spinless
         electrons residing on the two levels experience a
         repulsive interaction.
         (b) The mapping onto a spinful generalized Anderson
         model, with a tilted magnetic field and different
         tunnelling elements for spin-up and spin-down
         electrons.
         (c) The low-energy behavior of the generalized
         Anderson model is mapped onto an anisotropic Kondo
         model with a tilted magnetic field,
         $\vec{h}_{\text{tot}}$.
} \label{fig:models}
\end{figure*}

In this paper we study the low-energy behavior of a generic
model, depicted in Fig.~\ref{fig:models}a, which pertains
either to a single two-level quantum dot or to a double quantum
dot where each dot harbors only a single level. The spin
degeneracy of the electrons is assumed to be lifted by an
external magnetic field. Several variants of this model have
been studied intensely in recent years, in conjunction with a
plethora of phenomena, such as many-body resonances in the
spectral density,~\cite{Boese01} phase lapses in the
transmission phase,~\cite{Silva02,Golosov06} charge
oscillations,~\cite{Gefen04,Sindel05} and correlation-induced
resonances in the conductance~\cite{Meden06PRL,Karrasch06}.
Albeit being described by the same model, no clear linkage has
been established between these seemingly different effects. The
reason is in part due to the large number of model parameters
involved, which so far obscured a clear physical picture. While
some exact statements can be made, these are restricted to
certain solvable limits,~\cite{Boese01} and are apparently
nongeneric~\cite{Meden06PRL}. Here we construct a framework
which encompasses all parameter regimes of the model, and
enables a unified description of the various phenomena alluded
to above, exposing their common physical origin. For the most
interesting regime of strong fluctuations between the two
levels, we are able to give: (i)   explicit analytical
conditions for the
      occurrence of transmission phase lapses;
(ii)  an explanation of the population inversion and the
      charge oscillations~\cite{Gefen04,Sindel05,Silvestrov00}
      (including a Kondo enhancement of the latter);
(iii) a complete account of the correlation-induced
      resonances~\cite{Meden06PRL} as a disguised Kondo phenomenon.

After introducing the details of the double-dot Hamiltonian in
Sec.~\ref{sec:Model}, we begin our analysis by constructing a
linear transformation of the dot operators, \emph{and} a
simultaneous (generally different) linear transformation of the
lead operators, such that the 2$\times$2 tunnelling matrix
between the two levels on the dot and the leads becomes
diagonal (with generally different eigenvalues). As a result,
the electrons acquire a pseudo-spin degree of freedom which is
conserved upon tunnelling between the dot and the continuum, as
shown schematically in Fig.~\ref{fig:models}b. Concomitantly,
the transformation generates a local Zeeman magnetic field. In
this way the original double-dot model system is transformed
into a generalized Anderson impurity model in the presence of a
(generally tilted) external magnetic field. This first stage is
detailed in Sec.~\ref{sec:ModelAnderson} and
Appendix~\ref{App:SVDdetails}.

We next analyze in Sec.~\ref{sec:LocalMoment} the low-energy
properties of our generalized Anderson model. We confine
ourselves to the local moment regime, in which there is a
single electron on the impurity. The fluctuations of the
pseudo-spin degree of freedom (which translate into charge
fluctuations between the two localized levels in the original
model) are determined by two competing effects: the polarizing
effect of the local magnetic field, and the Kondo screening by
the itinerant electrons. In order to quantitatively analyze
this competition, we derive an effective low-energy Kondo
Hamiltonian, using Haldane's scaling
procedure,~\cite{HaldanePRL78} together with the
Schrieffer-Wolff~\cite{Wolff66} transformation and Anderson's
poor man's scaling~\cite{Anderson70}. This portion of the
derivation resembles recent studies of the Kondo effect in the
presence of ferromagnetic leads,~\cite{Martinek03PRL} although
the physical context and implications are quite different.

As is mentioned above, the tunnelling between the impurity and
the continuum in the generalized Anderson model is (pseudo)
spin dependent. This asymmetry results in two important
effects:
(a) different renormalizations of the two local levels,
    which in turn generates an additional local magnetic
    field~\cite{Martinek03PRL}. This field is not necessarily
    aligned with the original Zeeman field that is present
    in the generalized Anderson model.
(b) An anisotropy of the exchange coupling between the
    conduction electrons and the local moment in the Kondo
    Hamiltonian.
However, since the scaling equations for the anisotropic Kondo
model~\cite{Anderson70,AndersonYuvalHamann70} imply a flow
towards the \emph{isotropic} strong coupling fixed point, the
low-energy behavior of the generalized Anderson model can be
still described in terms of two competing energy scales, the
Kondo temperature, $T^{}_{K}$, and the renormalized magnetic
field, $h_{\text{tot}}$. Our two-stage mapping, double-dot
$\Rightarrow$ generalized Anderson model $\Rightarrow$
anisotropic Kondo model (see Fig.~\ref{fig:models}), allows us
to obtain analytic expressions for the original model
properties in terms of those of the Kondo model. We derive in
Sec.~\ref{sec:observables} the occupation numbers on the two
localized levels by employing the Bethe \emph{ansatz} solution
of the magnetization of a Kondo spin in a finite magnetic
field~\cite{AndreiRMP83,WiegmannA83}. This solution also
results in a highly accurate expression for the conductance
based upon the Friedel-Langreth sum rule~\cite{Langreth66}.
Perhaps most importantly, it provides a single coherent picture
for the host of phenomena to which our model has been applied.

Examples of explicit results stemming from our general analysis
are presented in Sec.~\ref{sec:results}. First, we consider the
case in which the tunnelling is isotropic, being the same for
spin-up and spin-down electrons. Then the model is exactly
solvable by direct application of the Bethe \emph{ansatz} to
the Anderson Hamiltonian~\cite{WiegmannC83,WiegmannA83}. We
solve the resulting equations~\cite{Okiji82,WiegmannC83}
numerically and obtain the occupation numbers for arbitrary
parameter values of the model, and in particular, for arbitrary
values of the local Zeeman field. By comparing with the
occupation numbers obtained in Sec.~\ref{sec:observables} from
the Kondo version of the model, we are able to test the
accuracy of the Schrieffer-Wolff mapping onto the Kondo
Hamiltonian. We find that this mapping yields extremely precise
results over the entire local-moment regime. This exactly
solvable example has another virtue. It clearly demonstrates
the competition between the Kondo screening of the local spin,
which is governed by $T_K$, and the polarizing effect of the
local field $h_{\text{tot}}$. This competition is reflected
in the charging process of the quantum dot described by the
original Hamiltonian. We next proceed to apply our general
method to the features for which the anisotropy in the
tunnelling is relevant, notably the transmission phase lapses
and the correlation-induced resonances~\cite{Meden06PRL}. In
particular, we derive analytical expressions for the occupation
numbers and the conductance employing the mapping onto the
Kondo Hamiltonian. These analytical expressions give results
which are in a very good agreement with the data presented by
Meden and Marquardt,~\cite{Meden06PRL} which was obtained by
the functional and numerical renormalization-group methods
applied to the original model.

As our treatment makes extensive usage of the exact Bethe
\emph{ansatz} solutions for the impurity magnetization in the
isotropic Kondo and Anderson models with a finite magnetic
field, all relevant details of the solutions are concisely
gathered for convenience in Appendix~\ref{app:Bethe}.

\section{The double-dot system as a generalized Anderson model}
\label{sec:secII}

\subsection{The model\label{sec:Model}}

We consider spinless electrons in a system of two distinct
energy levels (a `quantum dot'), labelled $i = 1, 2$, which are
connected by tunnelling to two leads, labelled $\alpha = L, R$.
This quantum dot is penetrated by a (constant) magnetic flux.
The total Hamiltonian of the system reads
\begin{eqnarray}
\mathcal{H} = \mathcal{H}_l + \mathcal{H}_d + \mathcal{H}_{ld} \, ,
\label{IHAM}
\end{eqnarray}
in which $\mathcal{H}_{l}$ is the Hamiltonian of the leads,
$\mathcal{H}_{d}$ is the Hamiltonian of the isolated dot, and
$\mathcal{H}_{ld}$ describes the coupling between the dot and
the leads. The system is portrayed schematically in
Fig.~\ref{fig:models}a.

Each of the leads is modelled by a continuum of noninteracting
energy levels lying within a band of width $2D$, with a
constant density of states $\rho$~\cite{Comm-on-equal-rho}. The
corresponding Hamiltonian is given by
\begin{eqnarray}
\mathcal{H}_l = \sum_{ k\alpha} \varepsilon^{}_{k}
             c_{k\alpha}^{\dagger} c^{}_{k\alpha} \, ,
\end{eqnarray}
where $c^{\dagger}_{ k\alpha}$ ($c^{}_{ k\alpha}$) creates
(annihilates) an electron of wave vector $k$ on lead $\alpha$.
The two leads are connected to two external reservoirs, held at
the same temperature $T$ and having different chemical
potentials, $\mu_L$ and $\mu_R$, respectively. We take the
limit $\mu_L  \!\! \to \!\! \mu_R = 0$ in considering
equilibrium properties and the linear conductance.

The isolated dot is described by the Hamiltonian
\begin{equation}
\mathcal{H}_d = \left [
                 \begin{array}{cc}
                    d^{\dagger}_{1} & d^{\dagger}_{2}
                 \end{array}
         \right ] \cdot \hat{\mathcal{E}}_d \cdot
         \left [
                 \begin{array}{c}
                    d_{1} \\ d_{2}
                 \end{array}
         \right ] + U \, n_{1} n_{2} \, ,
\label{HDOT}
\end{equation}
where
\begin{eqnarray}
\hat{\mathcal{E}}_{d} = \frac{1}{2}
    \left [
            \begin{array}{cc}
                2 \, \epsilon_0 + \Delta &
                b \, e^{i(\varphi_{L}-\varphi_{R})/2} \\
                b \, e^{-i(\varphi_{L}-\varphi_{R})/2} &
                2 \, \epsilon_0  -\Delta
            \end{array}
    \right ] \,  .
\label{HPS}
\end{eqnarray}
Here, $d^{\dagger}_{i}$ ($d^{}_{i}$) creates (annihilates) an
electron on the $i$th level, $n_i \equiv d^{\dagger}_i d^{}_i$
are the occupation-number operators (representing the local
charge), $U>0$ denotes the Coulomb repulsion between electrons
that occupy the two levels, $\epsilon_{0} \pm \Delta /2$ are
the (single-particle) energies on the levels, and $b/2$ is the
amplitude for tunnelling between them. The phases $\varphi_{L}$
and $\varphi_{R}$, respectively, represent the Aharonov-Bohm
fluxes (measured in units of the flux quantum $2 \pi \hbar c/e$)
in the left and in the right hopping loops, such that the total
flux in the two loops is $\varphi \equiv \varphi_{L} +
\varphi_{R}$ [see Fig.~\ref{fig:models}a].

Gauge invariance grants us the freedom to distribute the
Aharonov-Bohm phases among the inter-dot coupling $b$ and the
couplings between the dot levels and the leads. With the
convention of Eq.~(\ref{HPS}), the coupling between the quantum
dot and the leads is described by the Hamiltonian
\begin{eqnarray}
\mathcal{H}_{ld} = \sum_{k}
       \left [
               \begin{array}{cc}
                   c^{\dagger}_{kL} & c^{\dagger}_{kR}
               \end{array}
       \right ] \cdot \hat{A} \cdot
       \left [
               \begin{array}{c}
                   d_1 \\ d_2
               \end{array}
       \right ] + \text{H.c.} \, ,
\end{eqnarray}
where
\begin{eqnarray}
\hat{A} =
    \left [
            \begin{array}{cc}
                  a_{L1}e^{i\varphi /2} & a_{L2}\\
                  a_{R1} & a_{R2}e^{i\varphi /2}
            \end{array}
    \right ]\, , \quad
    \varphi = \varphi_{L} + \varphi_{R} \, .
\label{AA}
\end{eqnarray}
Here the real (possibly negative) coefficients $a_{\alpha i}$
are the tunnelling amplitudes for transferring an electron from
the level $i$ to lead $\alpha$. Note that the Hamiltonian
depends solely on the total Aharonov-Bohm flux $\varphi$ when
the interdot coupling $b$ vanishes. Also, the tunnelling matrix
$\hat{A}$ is assumed to be independent of the wave vector $k$.
This assumption considerably simplifies the analysis while
keeping the main physical picture intact.

\subsection{Mapping onto a generalized Anderson model}
\label{sec:ModelAnderson}

The analysis of the model defined in Sec.~\ref{sec:Model}
employs an {\it exact} mapping of the Hamiltonian of
Eq.~(\ref{IHAM}) onto a generalized Anderson Hamiltonian, which
pertains to a single-level quantum dot, coupled to a
spin-degenerate band of conduction electrons. We show in
Appendix~\ref{App:SVDdetails} that the model depicted in
Fig.~\ref{fig:models}a is fully described by the Hamiltonian
\begin{widetext}
\begin{equation}
\mathcal{H} = \sum_{k, \sigma}
            \varepsilon_k \, c_{k\sigma}^{\dagger} c^{}_{k\sigma}
     + \sum_{\sigma}
            \Bigl (
                    \epsilon_0 - \sigma \frac{h}{2} \cos \theta
            \Bigr )
                       n_{\sigma}
     - \bigl ( d_{\spinup}^{\dagger} d^{}_{\spindown} +
        d_{\spindown}^{\dagger} d^{}_{\spinup} \bigr ) \,
        \frac{h}{2} \, \sin \theta
     + U n_{\spinup} n_{\spindown}
     + \sum_{k, \sigma} V^{}_{\sigma}
       \Bigl (
               c_{k\sigma}^{\dagger} d^{}_{\sigma} + \text{H.c.}
       \Bigr ) \, ,
\label{eq:Hand}
\end{equation}
\end{widetext}
schematically sketched Fig.~\ref{fig:models}b, which
generalizes the original Anderson model~\cite{Anderson61} in
two aspects. Firstly, it allows for spin-dependent coupling
between the dot and the conduction band. A similar variant of
the Anderson model has recently attracted much theoretical and
experimental attention in connection with the Kondo effect for
ferromagnetic
leads~\cite{Martinek03PRL,MartinekNRGferro,Pasupathy04,MartinekPRB05,
Comment-on-FM}. Secondly, it allows for a Zeeman field whose
direction is inclined with respect to the ``anisotropy'' axis
$z$. For spin-independent tunnelling, one can easily realign
the field along the $z$ axis by a simple rotation of the
different operators about the $y$ axis. This is no longer the
case once $V_{\spinup} \neq V_{\spindown}$, which precludes the
use of some of the exact results available for the Anderson
model. As we show below, the main effect of spin-dependent
tunnelling is to modify the effective field seen by electrons
on  the dot, by renormalizing its $z$-component.

The derivation of Eq.~(\ref{eq:Hand}) is accomplished by a
transformation known as the singular-value
decomposition,~\cite{Golub96} which allows one to express the
tunnelling matrix $\hat{A}$ in the form
\begin{equation}
\hat{A} = R_l^{\dagger} \cdot
    \left [
            \begin{array}{cc}
                  V_{\spinup} & 0 \\
                  0 & V_{\spindown}
            \end{array}
    \right ] \cdot R^{}_d \, .
\end{equation}
Here $R_{l}$ and $R_{d}$ are unitary 2$\times$2 matrices, which
are used to independently rotate the lead and the dot operators
according to
\begin{eqnarray}
      \left [
               \begin{array}{c}
                   d_{\spinup} \\ d_{\spindown}
               \end{array}
      \right ] \equiv R_{d} \cdot
      \left [
               \begin{array}{c}
                   d_{1} \\ d_{2}
               \end{array}
      \right ]\, ,  \quad
      \left [
               \begin{array}{c}
                   c_{k\spinup} \\ c_{k\spindown}
               \end{array}
      \right ] \equiv R_{l} \cdot
      \left [
               \begin{array}{c}
                   c_{kL} \\ c_{kR}
               \end{array}
      \right ]\, .
\label{eq:SVDdef}
\end{eqnarray}
To make contact with the conventional Anderson impurity model,
we have labelled the linear combinations of the original
operators [defined through Eqs.~(\ref{eq:SVDdef})] by the
``spin'' index $\sigma = \spinup$ ($+1$) and $\sigma =
\spindown$ ($-1$).

The transformation (\ref{eq:SVDdef}) generalizes the one in
which the \emph{same} rotation $R$ is applied to both the dot
and the lead operators. It is needed in the present, more
general, case since the matrix $\hat{A}$ generically lacks an
orthogonal basis of eigenvectors. The matrices $R_d$ and $R_l$
can always be chosen uniquely (up to a common overall phase)
such that~\cite{Comment-on-uniqueness} (a) the tunnelling
between the dot and the continuum is
    diagonal in the spin basis (so that the tunnelling
    conserves the spin);
(b) the amplitudes $V_{\spinup} \ge V_{\spindown} \ge 0$
    are real; and
(c) the part of the Hamiltonian of Eq.~(\ref{eq:Hand})
    pertaining to the dot has only real matrix elements
    with $h \sin \theta \geq 0$.
The explicit expressions for the rotation matrices $R_d$
and $R_l$ as well as for the model parameters appearing
in Eq.~(\ref{eq:Hand}) in terms of those of the original
Hamiltonian are given in Appendix~\ref{App:SVDdetails}.

It should be emphasized that partial transformations involving
only one rotation matrix, either $R_d$ or $R_l$, have
previously been applied in this context (see, e.g.,
Refs.~~\onlinecite{Boese01} and~~\onlinecite{Glazman01}).
However, excluding special limits, both $R_d$ and $R_l$ are
required to expose the formal connection to the Anderson model.
A first step in this direction was recently taken by Golosov
and Gefen~\cite{Golosov06}, yet only on a restricted
manifold for the tunnelling amplitudes $a_{\alpha i}$. In the
following section we discuss in detail the low-energy physics
of the Hamiltonian of Eq.~(\ref{eq:Hand}), focusing on the
local-moment regime. Explicit results for the conductance and
the occupations of the levels are then presented in
Secs.~\ref{sec:observables} and \ref{sec:results}.

\section{The local-moment regime\label{sec:LocalMoment}}

There are two limits where the model of Eq.~(\ref{IHAM}) has an
exact solution:~\cite{Boese01} (i) when the spin-down state is
decoupled in Eq.~(\ref{eq:Hand}), i.e., when $V_{\spindown} =
h\sin \theta = 0$; (ii) when the coupling is isotropic, i.e.,
$V_{\spinup} = V_{\spindown}$. In the former case,
$n_{\downarrow}$ is conserved. The Hilbert space separates then
into two disconnected sectors with $n_{\spindown} = 0$ and
$n_{\spindown} = 1$. Within each sector, the Hamiltonian can be
diagonalized independently as a single-particle problem. In the
latter case, one can always align the magnetic field $h$ along
the $z$ axis by a simple rotation of the different operators
about the $y$ axis. The model of Eq.~(\ref{eq:Hand}) reduces
then to a conventional Anderson model in a magnetic field, for
which an exact Bethe {\em ansatz} solution is
available~\cite{WiegmannA83}. (This special case will be
analyzed in great detail in Sec.~\ref{sec:ResultsIsotropic}.)

In terms of the model parameters appearing in the original
Hamiltonian, the condition $V_{\spindown} = 0$ corresponds to
\begin{equation}
      |a_{L1} a_{R2}| = |a_{R1} a_{L2}|, \ \ \text{and}
      \ \ \varphi = \beta \!\!\!\! \mod 2\pi ,
\label{exact-a}
\end{equation}
whereas $V_{\spinup} = V_{\spindown} = V$ corresponds to
\begin{equation}
      |a_{L1}| = |a_{R2}| , \
      |a_{L2}| = |a_{R1}| , \
      \text{and} \
      \varphi = (\pi + \beta) \!\!\!\! \mod 2\pi \, .
\label{exact-b}
\end{equation}
Here
\begin{equation}
\beta = \left \{
        \begin{array}{cc}
              0 & \text{if} \ \ a_{L1} a_{L2} a_{R1} a_{R2} > 0 \\ \\
              \pi & {\rm if}\ \ a_{L1} a_{L2} a_{R1} a_{R2} < 0
        \end{array}
        \right .
\end{equation}
records the combined signs of the four coefficients
$a_{\alpha i}$~\cite{Comment-on-phi}.

Excluding the two cases mentioned above, no exact solutions to
the Hamiltonian of Eq.~(\ref{IHAM}) are known.  Nevertheless,
we shall argue below that the model displays generic low-energy
physics in the ``local-moment'' regime, corresponding to the
Kondo effect in a finite magnetic field. To this end we focus
hereafter on $\Gamma_{\spinup}, \Gamma_{\spindown}, h \ll
-\epsilon_0, U + \epsilon_0$, and derive an effective
low-energy Hamiltonian for general couplings. Here
$\Gamma_{\sigma} = \pi \rho V_{\sigma}^2$ is half the
tunnelling rate between the spin state $\sigma$ and the leads.

\subsection{Effective low-energy Hamiltonian}

As is mentioned above, when $V_{\uparrow} = V_{\downarrow}$ one
is left with a conventional Kondo effect in the presence of a
finite magnetic field.  Asymmetry in the couplings,
$V_{\uparrow} \neq V_{\downarrow}$, changes this situation in
three aspects. Firstly, the effective magnetic field seen by
electrons on the dot is modified, acquiring a renormalized
$z$-component. Secondly, the elimination of the charge
fluctuations by means of a Schrieffer-Wolff
transformation,~\cite{Wolff66} results in an anisotropic
spin-exchange interaction. Thirdly, a new interaction term is
produced, coupling the spin and the charge. Similar aspects
have been previously discussed in the context of the Kondo
effect in the presence of ferromagnetic
leads,~\cite{Martinek03PRL} where the source of the asymmetry
is the inequivalent density of states for conduction electrons
with opposite spin~\cite{Comment-on-FM}. Below we elaborate on
the emergence of these features in the present case.

Before turning to a detailed derivation of the effective
low-energy Hamiltonian, we briefly comment on the physical
origin of the modified magnetic field. As is well known, the
coupling to the continuum renormalizes the bare energy levels
of the dot. For $\Gamma_{\uparrow}, \Gamma_{\downarrow}, h \ll
-\epsilon_0, U + \epsilon_0$, these renormalizations can be
accurately estimated using second-order perturbation theory in
$V_{\sigma}$. For $V_{\uparrow} \neq V_{\downarrow}$, each of
the bare levels $\epsilon_{\sigma} = \epsilon_0 - \frac{1}{2}
\sigma h \cos \theta$ is shifted by a different amount, which
acts in effect as an excess magnetic field. Explicitly, for $T
= 0$ and $D \gg |\epsilon_0|, U$ one
obtains~\cite{Martinek03PRL,Silvestrov00}
\begin{equation}
\Delta h_z =
       \frac{\Gamma_{\uparrow} - \Gamma_{\downarrow}}{\pi}
       \ln \frac{\epsilon_0 + U}{|\epsilon_0|} \, .
\label{Delta-h_z}
\end{equation}
As $\epsilon_0$ is swept across $-U/2$, $\Delta h_z \propto
\Gamma_{\uparrow} - \Gamma_{\downarrow}$ changes sign. Had
$|\Gamma_{\uparrow} - \Gamma_{\downarrow}|$ exceeded $h$ this
would have dictated a sign-reversal of the $z$-component of the
combined field as $\epsilon_0$ is tuned across the
Coulomb-blockade valley. As originally noted by Silvestrov and
Imry,~\cite{Silvestrov00} this simple but insightful
observation underlies the population inversion discussed in
Refs.~\onlinecite{Gefen04,Sindel05}
and~\onlinecite{Silvestrov00} for a singly occupied dot. We
shall return to this important point in greater detail later
on.

A systematic derivation of the effective low-energy Hamiltonian
for $\Gamma_{\spinup}, \Gamma_{\spindown}, h \ll -\epsilon_0, U
+ \epsilon_0$ involves the combination of Anderson's poor-man's
scaling~\cite{Anderson70} and the Schrieffer-Wolff
transformation~\cite{Wolff66}. For $|\epsilon_0| \sim U +
\epsilon_0$, the elimination of high-energy excitations
proceeds in three steps. First Haldane's perturbative scaling
approach~\cite{HaldanePRL78} is applied to progressively reduce
the bandwidth from its bare value $D$ down to $D_{\rm SW} \sim
|\epsilon_0| \sim U + \epsilon_0$. Next a Schrieffer-Wolff
transformation is carried out to eliminate charge fluctuations
on the dot. At the conclusion of this second step one is left
with a generalized Kondo Hamiltonian [Eq.~(\ref{H-Kondo})
below], featuring an anisotropic spin-exchange interaction and
an additional interaction term that couples spin and charge.
The Kondo Hamiltonian also includes a finite magnetic field
whose direction is inclined with respect to the anisotropy axis
$z$. In the third and final stage, the Kondo Hamiltonian is
treated using Anderson's poor-man's scaling~\cite{Anderson70}
to expose its low-energy physics.

The above procedure is further complicated in the case where
$|\epsilon_0|$ and $U + \epsilon_0$ are well separated in
energy. This situation requires   two distinct Schrieffer-Wolff
transformations: one at $D_{\rm SW}^{\rm up} \sim \max \{
|\epsilon_0|, U + \epsilon_0\}$ and the other at $D_{\rm
SW}^{\rm down} \sim \min \{ |\epsilon_0|, U + \epsilon_0\}$.
Reduction of the bandwidth from $D_{\rm SW}^{\rm up}$ to
$D_{\rm SW}^{\rm down}$ is accomplished using yet another
(third) segment of the perturbative scaling. It turns out that
all possible orderings of $|\epsilon_0|$ and $U + \epsilon_0$
produce  the same Kondo Hamiltonian, provided that
$\Gamma_{\uparrow}$, $\Gamma_{\downarrow}$ and $h$ are
sufficiently small. To keep the discussion as concise as
possible, we therefore restrict the presentation to the case
$|\epsilon_0| \sim U + \epsilon_0$.

Consider first the energy window between $D$ and $D_{\rm SW}$,
which is treated using Haldane's perturbative
scaling~\cite{HaldanePRL78}. Suppose that the bandwidth has
already been lowered from its initial value $D$ to some value
$D' = D e^{-l}$ with $0 < l < \ln (D/D_{SW})$. Further reducing
the bandwidth to $D'(1 - \delta l)$ produces a renormalization
of each of the energies $\epsilon_{\uparrow}$,
$\epsilon_{\downarrow}$, and $U$. Specifically, the
$z$-component of the magnetic field, $h_z \equiv
\epsilon_{\downarrow} - \epsilon_{\uparrow}$, is found to obey
the scaling equation
\begin{equation}
\frac{d h_z}{d l} =
     \frac{\Gamma_{\uparrow} - \Gamma_{\downarrow}}{\pi}
     \left [
              \frac{1}{1 - e^{l} \epsilon_0/D} -
              \frac{1}{1 + e^{l} (U + \epsilon_0)/D}
     \right ] .
\label{dh_z-dl}
\end{equation}
Here we have  retained $\epsilon_0$ and $U + \epsilon_0$ in the
denominators, omitting  corrections which are higher-order in
$\Gamma_{\uparrow}$, $\Gamma_{\downarrow}$, and $h$ (these
include also the small renormalizations of $\epsilon_{\sigma}$
and $U$ that are accumulated in the course of the scaling). The
$x$-component of the field, $h_x = h \sin \theta$, remains
unchanged throughout the procedure. Upon reaching $D' = D_{\rm
SW}$, the renormalized field $h_z$ becomes
\begin{equation}
h_z^{\ast} = h \cos \theta +
     \frac{\Gamma_{\uparrow} - \Gamma_{\downarrow}}{\pi}
     \ln \frac{D_{\rm SW} + U + \epsilon_0}
                      {D_{\rm SW} - \epsilon_0} \,  ,
\end{equation}
where we have assumed $D \gg |\epsilon_0|, U$.

Once the scale $D_{\rm SW}$ is reached, charge fluctuations on
the dot are eliminated  via a Schrieffer-Wolff
transformation,~\cite{Wolff66} which generates among other
terms also further renormalizations of $\epsilon_{\sigma}$.
Neglecting $h$ in the course of the transformation, one arrives
at the following Kondo-type Hamiltonian,
\begin{eqnarray}
\mathcal{H}_K &=& \sum_{k, \sigma}
           \varepsilon^{}_k c_{k\sigma}^{\dagger} c^{}_{k\sigma}
           +  J_{\perp} \left ( S_x s_x + S_y s_y \right )
           +  J_z S_z s_z
\nonumber\\
         &+& v_{\rm sc} S^z \sum_{k, k', \sigma}\!\!
              :\! c^{\dagger}_{k \sigma} c^{}_{k' \sigma}\!:
            + \sum_{k, k', \sigma}\! (v_+ + \sigma v_-)\!
              :\! c^{\dagger}_{k \sigma} c^{}_{k' \sigma}\!:
\nonumber\\
         &-& \tilde{h}_z S_z - \tilde{h}_x S_x .
\label{H-Kondo}
\end{eqnarray}
Here we have represented the local moment on the dot by
the spin-$\frac{1}{2}$ operator
\begin{equation}
\vec{S} = \frac{1}{2} \sum_{\sigma, \sigma'}
          \vec{\tau}^{}_{\sigma \sigma'}
          d^{\dagger}_{\sigma} d^{}_{\sigma'}
\end{equation}
($\vec{\tau}$ being the Pauli matrices), while
\begin{equation}
\vec{s} = \frac{1}{2} \sum_{k, k'} \sum_{\sigma, \sigma'}
          \vec{\tau}^{}_{\sigma \sigma'}
          c^{\dagger}_{k \sigma} c^{}_{k' \sigma'}
\end{equation}
are the local conduction-electron spin densities. The symbol
$:\!c^{\dagger}_{k \sigma} c^{}_{k' \sigma}\!\!:\ = c^{\dagger}_{k
\sigma} c^{}_{k' \sigma} - \delta_{k, k'} \theta(-\epsilon_k)$
stands for normal ordering with respect to the filled Fermi sea.
The various couplings that appear in Eq.~(\ref{H-Kondo}) are given
by the explicit expressions
\begin{equation}
\rho J_{\perp} =
       \frac{2 \sqrt{\Gamma_{\uparrow} \Gamma_{\downarrow}}}
            {\pi}
       \left (
               \frac{1}{|\epsilon_0|}
               + \frac{1}{U + \epsilon_0}
       \right ) ,
\label{J-perp}
\end{equation}
\begin{equation}
\rho J_{z} =
       \frac{\Gamma_{\uparrow} + \Gamma_{\downarrow}}
            {\pi}
       \left (
               \frac{1}{|\epsilon_0|}
               + \frac{1}{U + \epsilon_0}
       \right ) ,
\label{J-z}
\end{equation}
\begin{equation}
\rho v_{\rm sc} =
       \frac{\Gamma_{\uparrow} - \Gamma_{\downarrow}}
              {4 \pi}
       \left (
               \frac{1}{|\epsilon_0|}
               + \frac{1}{U + \epsilon_0}
       \right ) ,
\end{equation}
\begin{equation}
\rho v_{\pm} =
       \frac{\Gamma_{\uparrow} \pm \Gamma_{\downarrow}}
            {4 \pi}
       \left (
               \frac{1}{|\epsilon_0|}
               - \frac{1}{U + \epsilon_0}
       \right ) ,
\label{v-pm}
\end{equation}
\begin{equation}
\tilde{h}_z = h \cos \theta +
            \frac{\Gamma_{\uparrow} - \Gamma_{\downarrow}}
                 {\pi}
            \ln  \frac{U + \epsilon_0}{|\epsilon_0|} \,  ,
\label{h_z-tilde}
\end{equation}
and
\begin{equation}
\tilde{h}_x = h \sin \theta .
\label{h_x-tilde}
\end{equation}

Equations~(\ref{J-perp})--(\ref{h_x-tilde}) are correct to
leading order in $\Gamma_{\uparrow}$, $\Gamma_{\downarrow}$,
and $h$, in accordance with the inequality $\Gamma_{\uparrow},
\Gamma_{\downarrow}, h \ll |\epsilon_0|, U + \epsilon_0$. In
fact, additional terms are generated in Eq.~(\ref{H-Kondo})
when  $h$ is kept in the course of the Schrieffer-Wolff
transformation. However, the neglected terms are smaller than
the ones retained by a factor of $h/\min \{|\epsilon_0|, U +
\epsilon_0 \} \ll 1$, and are not expected to alter the
low-energy physics in any significant way. We also note that
$\tilde{h}_z$ accurately reproduces the second-order correction
to $h_z$ detailed in Eq.~(\ref{Delta-h_z}). As emphasized
above, the same effective Hamiltonian is obtained when
$|\epsilon_0|$ and $U + \epsilon_0$ are well separated in
energy, although the derivation is notably more cumbersome. In
unifying the different possible orderings of $|\epsilon_0|$ and
$U + \epsilon_0$, the effective bandwidth in
Eq.~(\ref{H-Kondo}) must be taken to be $D_0 \sim \min
\{|\epsilon_0|, U + \epsilon_0\}$.

\subsection{Reduction to the Kondo effect in a finite
            magnetic field}

In addition to spin-exchange anisotropy and a tilted magnetic
field, the Hamiltonian of Eq.~(\ref{H-Kondo}) contains a new
interaction term, $v_{\rm sc}$, which couples spin and charge.
It also includes spin-dependent potential scattering,
represented by the term $v_{-}$ above. As is well known,
spin-exchange anisotropy is irrelevant for the conventional
spin-$\frac{1}{2}$ single-channel Kondo problem. As long as one
lies within the confines of the antiferromagnetic domain, the
system flows to the same strong-coupling fixed point no matter
how large the exchange anisotropy is. SU(2) spin symmetry is
thus restored at low energies. A finite magnetic field $h$ cuts
off the flow to isotropic couplings, as does the temperature
$T$. However, the residual anisotropy is negligibly small if
$h$, $T$ and the bare couplings are small. That is,
low-temperature thermodynamic and dynamic quantities follow a
single generic dependence on $T/T_K$ and $h/T_k$, where $T_K$
is the Kondo temperature. All relevant information on the bare
spin-exchange anisotropy is contained for weak couplings in the
microscopic form of $T_K$.

The above picture is insensitive to the presence of weak
potential scattering, which only slightly modifies the
conduction-electron phase shift at the Fermi energy. As we show
below, neither is it sensitive to the presence of the weak
couplings $v_{\rm sc}$ and $v_{-}$ in Eq.~(\ref{H-Kondo}). This
observation is central to our discussion, as it enables a very
accurate and complete description of the low-energy physics of
$\mathcal{H}_K$ in terms of the conventional Kondo model in a
finite magnetic field. Given the Kondo temperature $T_K$ and
the direction and magnitude of the renormalized field
pertaining to Eq.~(\ref{H-Kondo}), physical observables can be
extracted from the exact Bethe {\em ansatz} solution of the
conventional Kondo model. In this manner, one can accurately
compute the conductance and the occupation of the levels, as
demonstrated in Secs.~\ref{sec:observables} and
\ref{sec:results}.

To establish this important point, we apply poor-man's
scaling~\cite{Anderson70} to the Hamiltonian of
Eq.~(\ref{H-Kondo}). Of the different couplings that appear in
$\mathcal{H}_K$, only $J_z$, $J_{\perp}$, and $\tilde{h}_z$ are
renormalized at second order. Converting to the dimensionless
exchange couplings $\tilde{J}_z = \rho J_z$ and
$\tilde{J}_{\perp} = \rho J_{\perp}$, these are found to obey
the standard scaling
equations~\cite{AndersonYuvalHamann70,Anderson70}
\begin{eqnarray}
\frac{ d\tilde{J}_z }{dl} &=& \tilde{J}_{\perp}^2 \,  ,
\label{scaling-J_z} \\
\frac{ d\tilde{J}_{\perp} }{dl} &=&
            \tilde{J}_z \tilde{J}_{\perp} \, ,
\label{scaling-J_perp}
\end{eqnarray}
independent of $v_{\rm sc}$ and $v_{\pm}$. Indeed, the
couplings $v_{\rm sc}$ and $v_{\pm}$ do not affect the scaling
trajectories in any way, other than through a small
renormalization to $\tilde{h}_z$:
\begin{equation}
\frac{d \tilde{h}_z}{d l} =
        D_0 \, e^{-l}
        \left (
                  \tilde{J}_z \tilde{v}_{-} +
                  2 \tilde{v}_{\rm sc} \tilde{v}_{+}
        \right ) 8 \ln 2 .
\label{scaling-h_z}
\end{equation}
Here $\tilde{v}_{\mu}$ are the dimensionless couplings $\rho
v_{\mu}$ ($\mu = {\rm sc}, \pm$), and $l$ equals $\ln (D_0
/D')$ with $D'$ the running bandwidth.

As stated above, the scaling equations
(\ref{scaling-J_z})--(\ref{scaling-J_perp}) are identical to
those obtained for the conventional anisotropic Kondo model.
Hence, the Kondo couplings flow toward strong coupling along
the same scaling trajectories and with the same Kondo
temperature as in the absence of $v_{\rm sc}$ and $v_{\pm}$.
Straightforward integration of
Eqs.~(\ref{scaling-J_z})--(\ref{scaling-J_perp}) yields
\begin{equation}
T_K = D_0 \exp
      \left (
               -\frac{1}{\rho \, \xi} \tanh^{-1}\!
                    \frac{\xi}{J_{z}}
      \right )
\label{scaling-T_K-1}
\end{equation}
with $\xi = \sqrt{J_z^2 - J_{\perp}^2}$. Here we have exploited
the hierarchy $J_z \geq J_{\perp} > 0$ in deriving
Eq.~(\ref{scaling-T_K-1}). In terms of the original model
parameters appearing in Eq.~(\ref{eq:Hand}),
Eq.~(\ref{scaling-T_K-1}) takes the form
\begin{equation}
T_K = D_0 \exp
  \left [
           \frac{\pi \epsilon_0 (U + \epsilon_0)}
                {2U(\Gamma_{\uparrow}-\Gamma_{\downarrow})}
           \ln\!
           \frac{\Gamma_{\uparrow}}{\Gamma_{\downarrow}}
  \right ] \, .
\label{scaling-T_K-2}
\end{equation}
Equation~(\ref{scaling-T_K-2}) was obtained within second-order
scaling, which is known to overestimate the pre-exponential
factor that enters $T_K$. We shall not seek an improved
expression for $T_K$ encompassing all parameter regimes of
Eq.~(\ref{eq:Hand}). More accurate expressions will be given
for the particular cases of interest, see
Sec.~\ref{sec:results} below. Much of our discussion will not
depend, though, on the precise form of $T_K$. We shall only
assume it to be sufficiently small such that the renormalized
exchange couplings can be regarded isotropic starting at
energies well above $T_K$.

The other competing scale which enters the low-energy physics
is the fully renormalized magnetic field: $\vec{h}_{\text{tot}}
= h^x_{\text{tot}}\, \hat{x} + h^z_{\text{tot}}\, \hat{z}$.
While the transverse field $h^x_{\text{tot}}$ remains given by
$h \sin \theta$, the longitudinal field $h^z_{\text{tot}}$ is
obtained by integration of Eq.~(\ref{scaling-h_z}), subject to
the initial condition of Eq.~(\ref{h_z-tilde}). Since the
running coupling $\tilde{J}_z$ is a slowly varying function of
$l$ in the range where
Eqs.~(\ref{scaling-J_z})--(\ref{scaling-h_z}) apply, it can be
replaced for all practical purposes by its bare value in
Eq.~(\ref{scaling-h_z}). Straightforward integration of
Eq.~(\ref{scaling-h_z}) then yields
\begin{eqnarray}
h_{\text{tot}}^{z} &=&
   h \cos \theta +
          \frac{\Gamma_{\uparrow} - \Gamma_{\downarrow}}
               {\pi}
          \ln   \frac{U + \epsilon_0}{|\epsilon_0|}
\nonumber\\
   &+&    3 \ln(2) \, D_0
          \frac{\Gamma^2_{\uparrow}-\Gamma^2_{\downarrow}}
               {\pi^2} \times
          \frac{U (U + 2 \epsilon_0)}
               {(U + \epsilon_0)^2 \epsilon_0^2} \, ,
\label{h-total}
\end{eqnarray}
where we have used Eqs.~(\ref{J-z})--(\ref{v-pm}) for $J_z$,
$v_{\rm sc}$, and $v_{\pm}$. Note that the third term on the
right-hand side of Eq.~(\ref{h-total}) is generally much
smaller than the first two terms, and can typically be
neglected.

To conclude this section, we have shown that the Hamiltonian of
Eq.~(\ref{eq:Hand}), and thus that of Eq.~(\ref{IHAM}), is
equivalent at sufficiently low temperature and fields to the
ordinary \emph{isotropic} Kondo model with a tilted magnetic
field, provided that $\Gamma_{\uparrow}, \Gamma_{\downarrow}
\ll |\epsilon_0|, U + \epsilon_0$. The relevant Kondo
temperature is approximately given by
Eq.~(\ref{scaling-T_K-2}), while the components of
$\vec{h}_{\text{tot}} = h^x_{\text{tot}}\, \hat{x} +
h^z_{\text{tot}}\, \hat{z}$ are given by $h^x_{\text{tot}} = h
\sin \theta$ and Eq.~(\ref{h-total}).

\section{Physical observables
\label{sec:observables}}

Having established the intimate connection between the
generalized Anderson Hamiltonian, Eq.~(\ref{eq:Hand}), and the
standard Kondo model with a tilted magnetic field, we now
employ well-known results of the latter model in order to
obtain a unified picture for the conductance and the occupation
of the levels of our original model, Eq.~(\ref{IHAM}). The
analysis extends over a rather broad range of parameters. For
example, when $U + 2\epsilon_0 = 0$, then the sole requirement
for the applicability of our results is for $\sqrt{\Delta^2 +
b^2}$ to be small. The tunnelling matrix $\hat{A}$ can be
practically arbitrary as long as the system lies deep in the
local-moment regime. The further one departs from the middle of
the Coulomb-blockade valley the more restrictive the condition
on $\hat{A}$ becomes in order for $\vec{h}_{\rm tot}$ to stay
small. Still, our approach is applicable over a surprisingly
broad range of parameters, as demonstrated below. Unless stated
otherwise, our discussion is restricted to zero temperature.

\subsection{Conductance}

At zero temperature, a local Fermi liquid is formed in the
Kondo model. Only elastic scattering takes place at the Fermi
energy, characterized by the scattering phase shifts for the
two appropriate conduction-electron modes. For a finite
magnetic field $h$ in the $z$-direction, single-particle
scattering is diagonal in the spin index. The corresponding
phase shifts, $\delta_{\uparrow}(h)$ and
$\delta_{\downarrow}(h)$, are given by the Friedel-Langreth
sum rule,~\cite{Langreth66,Comment-on-Langreth}
$\delta_{\sigma}(h) = \pi \qav{n_{\sigma}}$, which when applied
to the local-moment regime takes the form
\begin{equation}
\delta_{\sigma}(h) = \frac{\pi}{2} + \sigma \pi M(h) \, .
\label{phase-shift-1}
\end{equation}
Here $M(h)$ is the spin magnetization, which
reduces~\cite{Comment-on-M_K} in the scaling
regime to a universal function of $h/T_K$,
\begin{equation}
\label{eq:MhUniversal}
M(h) = M_K(h/T_K) \, .
\end{equation}
Thus, Eq.~(\ref{phase-shift-1}) becomes
$\delta_{\sigma}(h) = \pi/2 + \sigma \pi M_K(h/T_K)$,
where $M_K(h/T_K)$ is given by Eq.~(\ref{eq:MKfullWiegmann})

To apply these results to the problem at hand, one first needs
to realign the tilted field along the $z$ axis. This is
achieved by a simple rotation of the different operators about
the $y$ axis. Writing the field $\vec{h}_{\text{tot}}$ in the
polar form
\begin{align}
\label{eq:htotExplicit}
\vec{h}_{\text{tot}} & \equiv h_{\text{tot}}
        \left (
                \sin \theta_h \hat{x} +
                \cos \theta_h \hat{z}
        \right ) \nonumber\\
        & \approx h \sin \theta \, \hat{x} +
        \left ( h \cos \theta +
        \frac{\Gamma_{\uparrow} - \Gamma_{\downarrow}}
             {\pi}
        \ln \frac{U + \epsilon_0}{|\epsilon_0|}
            \right ) \hat{z} \, ,
\end{align}
the lead and the dot operators are rotated according to
\begin{equation}
   \left [
            \begin{array}{c}
                \tilde{c}_{k\spinup} \\
                \tilde{c}_{k\spindown}
            \end{array}
   \right ] = R_{h} \cdot
   \left [
            \begin{array}{c}
                c_{k\spinup} \\ c_{k\spindown}
            \end{array}
   \right ] = R_{h} R_{l} \cdot
   \left [
            \begin{array}{c}
                c_{kL} \\ c_{kR}
            \end{array}
   \right ]
\end{equation}
and
\begin{equation}
   \left [
            \begin{array}{c}
                \tilde{d}_{\spinup} \\
                \tilde{d}_{\spindown}
            \end{array}
   \right ] = R_{h} \cdot
   \left [
            \begin{array}{c}
                d_{\spinup} \\ d_{\spindown}
            \end{array}
   \right ] = R_h R_{d} \cdot
   \left [
            \begin{array}{c}
                d_{1} \\ d_{2}
            \end{array}
   \right ]\ , \label{eq:RhRddef}
\end{equation}
with
\begin{equation}
R_{h} = e^{i (\theta_h/2) \tau_y} =
      \left [
              \begin{array}{cc}
                \ \    \cos (\theta_h/2) &\
                        \sin (\theta_h/2) \\
                   - \sin (\theta_h/2) &\
                        \cos (\theta_h/2)
              \end{array}
      \right ] \, . \label{eq:Rhdef}
\end{equation}
Here $R_{l}$ and $R_{d}$ are the unitary matrices used in
Eq.~(\ref{eq:SVDdef}) to independently rotate the lead and the
dot operators. Note that since $\sin \theta \ge 0 $, the range
of $\theta_h$ is $\theta_h \in [0; \pi]$.

The new dot and lead degrees of freedom have their spins
aligned either parallel ($\tilde{d}_{\uparrow}$ and
$\tilde{c}_{k \uparrow}$) or antiparallel
($\tilde{d}_{\downarrow}$ and $\tilde{c}_{k \downarrow}$) to
the field $\vec{h}_{\text{tot}}$. In this basis the
single-particle scattering matrix is diagonal,
\begin{equation}
\tilde{S} =
    - \left [
            \begin{array}{cc}
                 e^{i 2 \pi M_K(h_{\text{tot}}/T_K)} &
                      0 \\
                 0 &
                 e^{-i 2 \pi M_K(h_{\text{tot}}/T_K)}
            \end{array}
      \right ] \, .
\label{Scatt-mat-hat}
\end{equation}
The conversion back to the original basis set of left- and
right-lead electrons is straightforward,
\begin{equation}
S = R^{\dagger}_{l} R^{\dagger}_{h} \tilde{S}
    R_{h}^{} R_{l}^{} \equiv
    \left [
            \begin{array}{cc}
                 r & t' \\
                 t & r'
            \end{array}
      \right ] \, ,
\label{Scatt-mat}
\end{equation}
providing us with the zero-temperature conductance
$G = (e^2/2 \pi \hbar) |t|^2$.

Equations~(\ref{Scatt-mat-hat}) and (\ref{Scatt-mat})
were derived employing the mapping of Eq.~(\ref{IHAM})
onto an effective isotropic Kondo model with a tilted
magnetic field, in the $v_{\rm sc}, v_{\pm} \to 0$
limit. Within this framework, Eqs.~(\ref{Scatt-mat-hat})
and (\ref{Scatt-mat}) are exact in the scaling regime,
$T_K/D_0 \ll 1$. The extent to which these equations
are indeed valid can be appreciated by considering the special case
$h \sin \theta = 0$, for which there exists an exact (and
independent) solution for the scattering matrix $S$ in terms of the
dot ``magnetization'' $M = \langle n_{\uparrow} - n_{\downarrow}
\rangle/2$ [see Eq.~(\ref{Scatt-mat-Langreth}) below].  That
solution, which is based on the Friedel-Langreth sum
rule~\cite{Langreth66} applied directly to a spin-conserving
Anderson model, reproduces Eqs.~(\ref{Scatt-mat-hat}) and
(\ref{Scatt-mat}) in the Kondo regime.

\subsubsection{Zero Aharonov-Bohm fluxes}

Of particular interest is the case where no Aharonov-Bohm
fluxes are present, where further analytic progress can
be made. For $\varphi_L = \varphi_R = 0$, the parameters
that appear in the Hamiltonian of Eq.~(\ref{IHAM}) are
all real. Consequently, the rotation matrices $R_{d}$
and $R_{l}$ acquire the simplified forms given by
Eqs.~\eqref{R_d-no-AB} and \eqref{R_l-no-AB}
(see Appendix~\ref{App:SVDdetails} for details). Under
these circumstances, the matrix product $R_{h} R_{l}$
becomes $\pm e^{i \tau_y (\theta_h + s_R \, \theta_l)/2}
e^{i \pi \tau_z (1 - s_R)/4}$, and the elements of the
scattering matrix [see Eq.~(\ref{Scatt-mat})] are
\begin{align}
t = t' =& - i \sin[2\pi M_K(h_{\text{tot}}/T_K)]
               \sin (\theta_l + s_R \theta_h) \ ,
\nonumber\\
r = (r')^{\ast} =&
          - \cos[2\pi M_K(h_{\text{tot}}/T_K)]
\nonumber\\
         &- i \sin[2\pi M_K(h_{\text{tot}}/T_K)]
               \cos (\theta_l + s_R \theta_h) \ .
\end{align}
Hence, the conductance is
\begin{align}
G = \frac{e^2}{2 \pi \hbar}
    \sin^2[2\pi M_K(h_{\text{tot}}/T_K)]
    \sin^2(\theta_l + s_R \theta_h) \, ,
\label{G-no-flux}
\end{align}
where the sign $s_R$ and angle $\theta_l$ are given by
Eqs.~\eqref{eq:sR} and \eqref{eq:theta-no-AB}, respectively.
All dependencies of the conductance on the original model
parameters that enter Eq.~(\ref{IHAM}) are combined in
Eq.~(\ref{G-no-flux}) into two variables alone, $\theta_l + s_R
\theta_h$ and the reduced field $h_{\text{tot}}/T_K$. In
particular, $\theta_l$ is determined exclusively by the
tunnelling matrix $\hat{A}$, while $s_R$ depends additionally
on the two dot parameters $\Delta$ and $b$.

The conditions for a phase lapse to occur are particularly
transparent from Eq.~(\ref{G-no-flux}). These lapses correspond
to zeroes of $t$, and, in turn, of the conductance. There are
two possibilities for $G$ to vanish: either $h_{\text{tot}}$ is
zero, or $\theta_l + s_R \theta_h$ equals an integer multiple
of $\pi$. For example, when the Hamiltonian of
Eq.~(\ref{eq:Hand}) is invariant under the particle-hole
transformation $d_{\sigma} \to d_{\sigma}^{\dagger}$ and $c_{k
\sigma} \to -c_{k \sigma}^{\dagger}$ (which happens to be the
case whenever $\sqrt{ \Delta^2 + b^2} = 0$ and $U + 2\epsilon_0
= 0$), then $h_{\text{tot}}$ vanishes, and consequently the
conductance vanishes as well. A detailed discussion of the
ramifications of Eq.~(\ref{G-no-flux}) is held in
Sec.~\ref{sec:ResultsAnisotropic} below.

\subsubsection{Parallel-field configuration}

For $h \sin \theta = 0$, spin is conserved by the
Hamiltonian of Eq.~(\ref{eq:Hand}). We refer to this
case as the ``parallel-field'' configuration, since the
magnetic field is aligned with the anisotropy axis $z$.
For a parallel field, one can easily generalize the
Friedel-Langreth sum rule~\cite{Langreth66} to the
Hamiltonian of Eq.~(\ref{eq:Hand}).~\cite{MartinekNRGferro}
Apart from the need to consider
each spin orientation separately, details of the
derivation are identical to those for the ordinary
Anderson model,~\cite{Langreth66} and so is the
formal result for the $T = 0$ scattering phase
shift: $\delta_{\sigma} = \pi \Delta N_{\sigma}$,
where $\Delta N_{\sigma}$ is the number of
displaced electrons in the spin channel $\sigma$.
In the wide-band limit, adopted throughout our
discussion, $\Delta N_{\sigma}$ reduces to the
occupancy of the corresponding dot level,
$\langle n_{\sigma} \rangle$. The exact
single-particle scattering matrix then becomes
\begin{equation}
S = e^{i\pi \langle n_{\uparrow} + n_{\downarrow} \rangle}
    R^{\dagger}_{l} \cdot
    \left [
            \begin{array}{cc}
                 e^{i 2 \pi M} & 0 \\
                 0 & e^{-i 2 \pi M}
            \end{array}
      \right ] \cdot R_{l} \, ,
\label{Scatt-mat-Langreth}
\end{equation}
where $M = \langle n_{\uparrow}-n_{\downarrow} \rangle/2$
is the dot ``magnetization.''

Equation~(\ref{Scatt-mat-Langreth}) is quite general. It covers
all physical regimes of the dot, whether empty, singly occupied
or doubly occupied, and extends to arbitrary fluxes $\varphi_L$
and $\varphi_R$. Although formally exact, it does not specify
how the dot ``magnetization'' $M$ and the total dot occupancy
$\langle n_{\uparrow} + n_{\downarrow} \rangle$ relate to the
microscopic model parameters that appear in
Eq.~(\ref{eq:Hand}). Such information requires an explicit
solution for these quantities. In the Kondo regime considered
above, $\langle n_{\uparrow} + n_{\downarrow} \rangle$ is
reduced to one and $M$ is replaced by $\pm
M_K(h_{\text{tot}}/T_K)$. Here the sign depends on whether the
field $\vec{h}_{\text{tot}}$ is parallel or antiparallel to the
$z$ axis (recall that $h_{\text{tot}} \ge 0$ by definition). As
a result, Eq.~(\ref{Scatt-mat-Langreth}) reproduces
Eqs.~(\ref{Scatt-mat-hat})--(\ref{Scatt-mat}).

To carry out the rotation in Eq.~(\ref{Scatt-mat-Langreth}), we
rewrite it in the form
\begin{equation}
S = e^{i\pi \langle n_{\uparrow} + n_{\downarrow} \rangle}
    R^{\dagger}_{l}
            \left [
                    \cos (2 \pi M) + i\sin (2 \pi M) \tau_z
            \right ]
    R_{l} \, .
\end{equation}
Using the general form of Eq.~(\ref{eq:phasel}) for the
rotation matrix $R_{l}$, the single-particle scattering matrix
is written as $S = e^{i\pi \langle n_{\uparrow} +
n_{\downarrow} \rangle} \bar{S}$, where
\begin{eqnarray}
\bar{S} &=& \cos (2 \pi M)
            + i \sin (2 \pi M) \cos \theta_l\, \tau_z
\nonumber \\
  &+&
     i \sin (2 \pi M) \sin \theta_l
       \left [
               \cos \phi_l\, \tau_x + \sin \phi_l\, \tau_y
       \right ] \, .
\end{eqnarray}
The zero-temperature conductance,
$G = (e^2/2 \pi \hbar)|t|^2$, takes then the exact form
\begin{equation}
G = \frac{e^2}{2 \pi \hbar}
    \sin^2(2\pi M) \sin^2 \theta_l \, .
\label{G-parallel-field}
\end{equation}

Two distinct properties of the conductance are apparent form
Eq.~(\ref{G-parallel-field}). Firstly, $G$ is bounded by
$\sin^2 \theta_l $ times the conductance quantum unit $e^2/2
\pi \hbar$. Unless $\theta_{l}$ happens to equal $\pm \pi/2$,
the maximal conductance is smaller than $e^2/2 \pi \hbar$.
Secondly, $G$ vanishes for $M = 0$ and is maximal for $M = \pm
1/4$. Consequently, when $M$ is tuned from $M \approx -1/2$ to
$M \approx 1/2$ by varying an appropriate control parameter
(for example, $\epsilon_0$ when $\Gamma_{\uparrow} \gg
\Gamma_{\downarrow}$), then $G$ is peaked at the points where
$M = \pm 1/4$. In the Kondo regime, when $M \to \pm
M_K(h_{\text{tot}}/T_K)$, this condition is satisfied for
$h_{\text{tot}} \approx 2.4 T_K$. As we show in
Sec.~\ref{sec:ResultsAnisotropic}, this is the physical origin
of the correlation-induced peaks reported by Meden and
Marquardt.~\cite{Meden06PRL} Note that for a given fixed
tunnelling matrix $\hat{A}$ in the parallel-field
configuration, the condition for a phase lapse to occur is
simply for $M$ to vanish.

\subsection{Occupation of the dot levels}

Similar to the zero-temperature conductance, one can exploit
exact results of the standard Kondo model to obtain the
occupation of the levels at low temperatures and fields.
Defining the two reduced density matrices
\begin{equation}
O_d =
           \begin{bmatrix}
             \langle d^{\dagger}_1 d^{}_1 \rangle &
             \langle d^{\dagger}_2 d^{}_1 \rangle \\
             \langle d^{\dagger}_1 d^{}_2 \rangle &
             \langle d^{\dagger}_2 d^{}_2 \rangle
       \end{bmatrix}
\end{equation}
and
\begin{equation}
\tilde{O}_d =
       \begin{bmatrix}
             \langle
                     \tilde{d}^{\dagger}_{\uparrow}
                     \tilde{d}^{}_{\uparrow}
             \rangle &
             \langle
                     \tilde{d}^{\dagger}_{\downarrow}
                     \tilde{d}^{}_{\uparrow}
             \rangle \\
             \langle
                     \tilde{d}^{\dagger}_{\uparrow}
                     \tilde{d}^{}_{\downarrow}
             \rangle &
             \langle
                     \tilde{d}^{\dagger}_{\downarrow}
                     \tilde{d}^{}_{\downarrow}
             \rangle
       \end{bmatrix}
\, ,
\end{equation}
these are related through
\begin{equation}
O_d^{} = R^{\dagger}_{d} R^{\dagger}_{h} \tilde{O}_d^{}
      R^{}_{h} R^{}_{d} \, .
\label{O_d-via-t-O_d}
\end{equation}
Here $R_{h} R_{d}$ is the overall rotation matrix pertaining to
the dot degrees of freedom, see Eq.~\eqref{eq:RhRddef}.

At low temperatures, the mapping onto an isotropic Kondo model
implies
\begin{equation}
\tilde{O}_d  = \begin{bmatrix}
               \qav{\tilde{n}_{\spinup}} & 0  \\
                0 & \qav{\tilde{n}_{\spindown}} \\
               \end{bmatrix} \, ,
\label{t-O_d}
\end{equation}
where
\begin{equation}
\qav{\tilde{n}_{\sigma}} = n_{\text{tot}}/2 +
     \sigma \tilde{M} \, .
\label{eq:ntildeseparated}
\end{equation}
Here we have formally separated the occupancies
$\qav{\tilde{n}_{\sigma}}$ into the sum of a spin component and
a charge component. The spin component involves the
magnetization $\tilde{M}$ along the direction of the total
effective field $\vec{h}_{\text{tot}}$. The latter is well
described by the universal magnetization curve
$M_K(h_{\text{tot}}/T_K)$ of the Kondo model [see
Eq.~\eqref{eq:MKfullWiegmann}]. As for the total dot occupancy
$n_{\text{tot}}$, deep in the local-moment regime charge
fluctuations are mostly quenched at low temperatures, resulting
in the near integer valance $n_{\text{tot}} \approx 1$. One can
slightly improve on this estimate of $n_{\text{tot}}$ by
resorting to first-order perturbation theory in
$\Gamma_{\sigma}$ (and zeroth order in $h$):
\begin{align}
n_{\text{tot}} & \approx
        1 + \frac{\Gamma_{\spinup} +\Gamma_{\spindown}}{2 \pi}
       \left (
               \frac{1}{\epsilon_0}+\frac{1}{U+\epsilon_0}
       \right )
       = 1 -2 \rho v_{+} \, .
\label{eq:n0PT}
\end{align}
This low-order process does not enter the Kondo
effect, and is not contained in
$M_K(h_{\text{tot}}/T_K)$.~\cite{Comment-on-charge-fluc}
With the above approximations, the combination of
Eqs.~(\ref{O_d-via-t-O_d}) and (\ref{t-O_d}) yields
a general formula for the reduced density matrix
\begin{equation}
O_d = n_{\text{tot}}/2  + M_K(h_{\rm tot}/T_K)
                    R^{\dagger}_{d} R^{\dagger}_{h}
                    \tau_z R^{}_{h} R^{}_{d} \, .
\label{O_d-general}
\end{equation}

\subsubsection{Zero Aharonov-Bohm fluxes}

As in the case of the conductance, Eq.~(\ref{O_d-general})
considerably simplifies in the absence of Aharonov-Bohm
fluxes, when the combined rotation $R_{h} R_{d}$ equals
$(s_R s_{\theta})^{1/2} e^{i \tau_y (\theta_h +
s_{\theta} \theta_d)/2} e^{i \pi \tau_z (1 - s_{\theta})/4}$
[see Eqs.~\eqref{eq:Rhdef} and \eqref{R_d-no-AB}].
Explicitly, Eq.~(\ref{O_d-general}) becomes
\begin{eqnarray}
O_d = n_{\text{tot}}/2 &+&
        M_K(h_{\rm tot}/T_K)
           \cos (\theta_d + s_{\theta} \theta_h) \tau_z
\nonumber \\
    &+& M_K(h_{\rm tot}/T_K)
            \sin (\theta_d + s_{\theta} \theta_h) \tau_x \, ,
\label{O_d-no-flux}
\end{eqnarray}
where the sign $s_{\theta}$ and angle $\theta_d$ are given
by Eqs.~\eqref{eq:stheta} and \eqref{eq:theta-no-AB},
respectively.

Several observations are apparent from Eq.~(\ref{O_d-no-flux}).
Firstly, when written in the original ``spin'' basis
$d^{\dagger}_1$ and $d^{\dagger}_2$, the reduced density matrix
$O_d$ contains the off-diagonal matrix element $M_K(h_{\rm
tot}/T_K) \sin (\theta_d + s_{\theta} \theta_h)$. The latter
reflects the fact that the original ``spin'' states are
inclined with respect to the anisotropy axis dynamically
selected by the system. Secondly, similar to the conductance of
Eq.~(\ref{G-no-flux}), $O_{d}$ depends on two variables alone:
$\theta_d + s_{\theta} \theta_h$ and the reduced field
$h_{\text{tot}}/T_K$. Here, again, the angle $\theta_d$ depends
solely on the tunnelling matrix $\hat{A}$, while the sign
$s_{\theta}$ depends additionally on $\Delta$ and $b$. Thirdly,
the original levels $d^{\dagger}_1$ and $d^{\dagger}_2$ have
the occupation numbers
\begin{subequations}
\label{eq:Actualn1n2}
\begin{align}
\qav{n_1} &=
       n_{\text{tot}}/2 + M_K(h_{\text{tot}}/T_K)
       \cos (\theta_d + s_{\theta} \theta_h) \, , \\
\qav{n_2} &=
       n_{\text{tot}}/2 - M_K(h_{\text{tot}}/T_K)
       \cos (\theta_d + s_{\theta} \theta_h) \, .
\end{align}
\end{subequations}
In particular, equal populations $\langle n_1 \rangle = \langle
n_2 \rangle$ are found if either $h_{\text{tot}}$ is zero or if
$\theta_d +s_{\theta} \theta_d$ equals $\pi/2$ up to an integer
multiple of $\pi$. This provides one with a clear criterion for
the occurrence of population
inversion,~\cite{Gefen04,Sindel05,Silvestrov00} i.e., the
crossover from $\qav{n_1} > \qav{ n_2}$ to
$\qav{n_2} > \qav{n_1}$ or vice versa.

\subsubsection{Parallel-field configuration}
\label{sec:occupany-PF}

In the parallel-field configuration, the angle $\theta_h$ is
either zero or $\pi$, depending on whether the magnetic field
$\vec{h}_{\text{tot}}$ is parallel or antiparallel to the $z$
axis (recall that $h \sin \theta = h_{\text{tot}} \sin
\theta_h=0$ in this case). The occupancies $\langle n_1
\rangle$ and $\langle n_2 \rangle$ acquire the exact
representation
\begin{subequations}
\label{eq:Actualn1n2-PF}
\begin{align}
\langle n_1 \rangle &=
       n_{\text{tot}}/2 + M \cos \theta_d \, , \\
\langle n_2 \rangle &=
       n_{\text{tot}}/2 - M \cos \theta_d \, ,
\end{align}
\end{subequations}
where $n_{\text{tot}}$ is the exact total occupancy of the dot
and $M = \langle n_{\uparrow} - n_{\downarrow} \rangle/2$ is
the dot ``magnetization,'' defined and used previously (not to
be confused with $\tilde{M} = \pm M$). As with the conductance,
Eqs.~(\ref{eq:Actualn1n2-PF}) encompass all regimes of the dot,
and extend to arbitrary Aharonov-Bohm fluxes. They properly
reduce to Eqs.~(\ref{eq:Actualn1n2}) in the Kondo regime, when
$n_{\text{tot}} \approx 1$ [see Eq.~\eqref{eq:n0PT}] and $M \to
\pm M_K(h_{\text{tot}}/T_K)$. [Note that Eqs.~(\ref{eq:Actualn1n2})
have been derived for zero Aharonov-Bohm fluxes.]

One particularly revealing observation that follows from
Eqs.~(\ref{eq:Actualn1n2-PF}) concerns the connection between
the phenomena of population inversion and phase lapses in the
parallel-field configuration. For a given fixed tunnelling
matrix $\hat{A}$ in the parallel-field configuration, the
condition for a population inversion to occur is identical to
the condition for a phase lapse to occur. Both require that $M
= 0$. Thus, these seemingly unrelated phenomena are synonymous
in the parallel-field configuration. This is not generically
the case when $h_{\text{tot}}^x\neq 0$, as can be seen, for
example, from Eqs.~(\ref{G-no-flux}) and (\ref{eq:Actualn1n2}).
In the absence of Aharonov-Bohm fluxes, the conductance is
proportional to $\sin^2(\theta_l + s_R \theta_h)$. It therefore
vanishes for $h_{\text{tot}}^x\neq 0$ only if $\theta_l + s_R
\theta_h = 0\!\!\mod\!\pi$. By contrast, the difference in
populations $\langle n_1 - n_2 \rangle$ involves the unrelated
factor $\cos (\theta_d + s_{\theta}\theta_h)$, which generally
does not vanish together with $\sin(\theta_l + s_R \theta_h)$.

Another useful result which applies to the parallel-field
configuration is an exact expression for the $T = 0$
conductance in terms of the population difference $\langle n_1
- n_2 \rangle$. It follows from Eqs.~(\ref{eq:Actualn1n2-PF})
that $M = \qav{n_1 - n_2}/( 2 \cos \theta_d)$. Inserting this
relation into Eq.~(\ref{G-parallel-field}) yields
\begin{align}
G = \frac{e^2}{2 \pi \hbar}
    \sin^2 \left (
                    \frac{\pi \langle n_1-n_2 \rangle}
                         {\cos \theta_d}
           \right )
    \sin^2 \theta_l \, .
\label{G-parallel}
\end{align}
This expression will be used in Sec.~\ref{sec:results} for
analyzing the conductance in the presence of isotropic
couplings, and for the cases considered by Meden and
Marquardt~\cite{Meden06PRL}.

\section{Results}
\label{sec:results}

Up until this point we have developed a general framework for
describing the local-moment regime in terms of two competing
energy scales, the Kondo temperature $T_K$ and the renormalized
magnetic field $h_{\text{tot}}$. We now turn to explicit
calculations that exemplify these ideas. To this end, we begin
in Sec.~\ref{sec:ResultsIsotropic} with the exactly solvable
case $V_{\spinup} = V_{\spindown}$, which corresponds to the
conventional Anderson model in a finite magnetic
field~\cite{Boese01}. Using the exact Bethe \emph{ansatz}
solution of the Anderson model,~\cite{WiegmannA83} we present a
detailed analysis of this special case with three objectives in
mind: (i) to benchmark our general treatment against
    rigorous results;
(ii) to follow in great detail the delicate interplay
     between the two competing energy scales that
     govern the low-energy physics;
(iii) to set the stage for the complete explanation of the
      charge oscillations~\cite{Gefen04,Sindel05,Silvestrov00}
      and the correlation-induced resonances in the
      conductance of this device~\cite{Meden06PRL,Karrasch06}.

We then proceed in Sec.~\ref{sec:ResultsAnisotropic} to the
generic anisotropic case $V_{\spinup} \neq V_{\spindown}$. Here
a coherent explanation is provided for the ubiquitous phase
lapses,~\cite{Golosov06} population
inversion,~\cite{Gefen04,Sindel05} and correlation-induced
resonances~\cite{Meden06PRL,Karrasch06} that were reported
recently in various studies of two-level quantum dots. In
particular, we expose the latter resonances as a disguised
Kondo phenomenon. The general formulae of
Sec.~\ref{sec:observables} are quantitatively compared to the
numerical results of Ref.~\onlinecite{Meden06PRL}. The detailed
agreement that is obtained nicely illustrates the power of the
analytical approach put forward in this paper.

\subsection{Exact treatment of $V_{\spinup}=V_{\spindown}$}
\label{sec:ResultsIsotropic}

As emphasized in Sec.~\ref{sec:LocalMoment}, all tunnelling
matrices $\hat{A}$ which satisfy Eq.~\eqref{exact-b} give rise
to equal amplitudes $V_{\spinup} = V_{\spindown} = V$ within
the Anderson Hamiltonian description of Eq.~\eqref{eq:Hand}.
Given this extra symmetry, one can always choose the unitary
matrices $R_{l}$ and $R_{d}$ in such a way that the magnetic
field $h$ points along the $z$ direction [namely, $\cos \theta
= 1$ in Eq.~\eqref{eq:Hand}]. Perhaps the simplest member in
this class of tunnelling matrices is the case where $a_{L1} =
-a_{L2} = a_{R1} = a_{R2} = V/\sqrt{2}$, $\varphi_L = \varphi_R
= 0$ and $b=0$. One can simply convert the conduction-electron
operators to even and odd combinations of the two leads,
corresponding to choosing $\theta_l = \pi/2 + \theta_d$.
Depending on the sign of $\Delta$, the angle $\theta_d$ is
either zero (for $\Delta < 0$) or $\pi$ (for $\Delta > 0$),
which leaves us with a conventional Anderson impurity in the
presence of the magnetic field $\vec{h} = |\Delta| \, \hat{z}$.
All other rotation angle that appear in Eqs.~\eqref{eq:phased}
and \eqref{eq:phased} (i.e., $\chi$'s and $\phi$'s) are equal
to zero. For concreteness we shall focus hereafter on this
particular case, which represents, up to a simple rotation of
the $d^{\dagger}_{\sigma}$ and $c^{\dagger}_{k \sigma}$
operators, all tunnelling matrices $\hat{A}$ in this category
of interest. Our discussion is restricted to zero temperature.

\subsubsection{Impurity magnetization}
\label{sec:ResTests}

\begin{figure}[t]
\includegraphics[width=7cm]{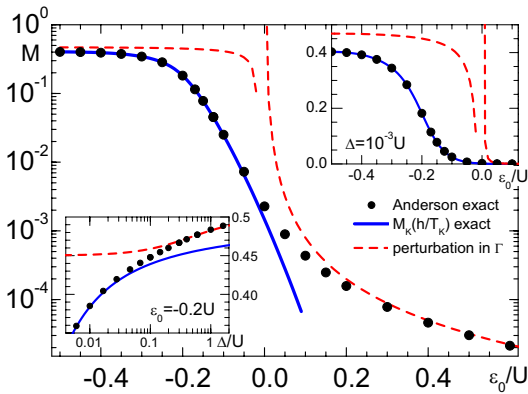}
\caption{(Color online) Magnetization of the isotropic
         case as a function of $\epsilon_0$: exact Bethe
         \emph{ansatz} curve and comparison with different
         approximation schemes.
         Black symbols show the magnetization $M$
         derived from the exact Bethe \emph{ansatz}
         equations; the dashed (red) line marks the result
         of first-order perturbation theory in $\Gamma$
         (Ref.~\onlinecite{Gefen04}, divergent at
         $\epsilon_0 = 0$); the thick (blue) line is the
         analytical formula for the magnetization in the
         Kondo limit, Eq.~\eqref{eq:MKfullWiegmann}, with
         $T_K$ given by Eq.~\eqref{eq:TKAndersonAccurate}.
         The model parameters are $\Gamma/U = 0.05$,
         $\Delta/U = 10^{-3}$ and $T = 0$.
         The upper right inset shows the same data but
         on a linear scale.
         The lower left inset shows the magnetization $M$
         as a function of the magnetic field $h = \Delta$
         at fixed $\epsilon_0/U = -0.2$. The universal
         magnetization curve of the Kondo model well
         describes the exact magnetization up to
         $M \approx 0.42$ (lower fields not shown),
         while first-order perturbation theory in
         $\Gamma$ fails from $M \approx 0.46$ downwards.}
\label{fig:MethodCompare}
\end{figure}

We have solved the exact Bethe \emph{anstaz} equations
numerically using the procedure outlined in
Appendix~\ref{app:Bethe}. Our results for the occupation
numbers $\qav{n_{\sigma}}$ and the magnetization $M =
\qav{n_{\uparrow} - n_{\downarrow}}/2$ are summarized in
Figs.~\ref{fig:MethodCompare} and \ref{fig:LargeFeature}.
Figure~\ref{fig:MethodCompare} shows the magnetization of the
Anderson impurity as a function of the (average) level position
$\epsilon_0$ in a constant magnetic field, $h = \Delta =
10^{-3} U$. The complementary regime $\epsilon_0 < -U/2$ is
obtained by a simple reflection about $\epsilon_0 = -U/2$, as
follows from the particle-hole transformation $d_{\sigma} \to
d_{-\sigma}^{\dagger}$ and $c_{k \sigma} \to -c_{k
-\sigma}^{\dagger}$. The Bethe \emph{ansatz} curve accurately
crosses over from the perturbative domain at large $\epsilon_0
\gg \Gamma$ (when the dot is almost empty) to the local-moment
regime with a fully pronounced Kondo effect (when the dot is
singly occupied). In the latter regime, we find excellent
agreement with the analytical magnetization curve of the Kondo
model, Eq.~\eqref{eq:MKfullWiegmann}, both as a function of
$\epsilon_0$ and as a function of the magnetic field $\Delta$
(lower left inset to Fig.~\ref{fig:MethodCompare}). The
agreement with the universal Kondo curve is in fact quite
surprising in that it extends nearly into the mixed-valent
regime. As a function of field, the Kondo curve of
Eq.~(\ref{eq:MKfullWiegmann}) applies up to fields of the order
of $h \sim \sqrt{\Gamma U} \gg T_K$.

\subsubsection{Occupation numbers and charge oscillations}
\label{sec:ResCharging}

\begin{figure}[t]
\includegraphics[width=7.5cm]{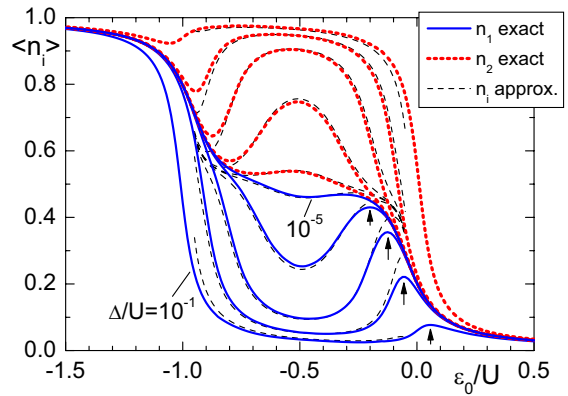}
\caption{(Color online) The occupation numbers $\qav{n_1}$
         [solid (blue) lines] and $\qav{n_2}$ [dotted (red)
         lines] versus $\epsilon_0$, as obtained from
         the solution of the exact Bethe \emph{ansatz}
         equations. In going from the inner-most to the
         outer-most pairs of curves, the magnetic field
         $h = \Delta$ increases by a factor of $10$
         between each successive pair of curves, with
         the inner-most (outer-most) curves corresponding
         to $\Delta/U = 10^{-5}$ ($\Delta/U = 0.1$).
         The remaining model parameters are
         $\Gamma/U = 0.05$ and $T=0$.
         Nonmonotonicities are seen in the process
         of charging. These are most pronounced for
         intermediate values of the field. The evolution
         of the nonmonotonicities with increasing field
         is tracked by arrows. The dashed black lines
         show the approximate values calculated from
         Eqs.~\eqref{eq:Actualn1n2} and (\ref{eq:n0PT})
         based on the mapping onto the Kondo Hamiltonian
         (here $\theta_h = 0$ and $\theta_d = \pi$).}
\label{fig:LargeFeature}
\end{figure}

Figure~\ref{fig:LargeFeature} displays the individual
occupation numbers $\qav{n_1}$ and $\qav{n_2}$ as a function of
$\epsilon_0$, for a series of constant fields $h = \Delta$. In
going from large $\epsilon_0 \gg \Gamma$ to large $-(\epsilon_0
+ U) \gg \Gamma$, the total charge of the quantum dot increases
monotonically from nearly zero to nearly two. However, the
partial occupancies $\qav{n_1}$ and $\qav{n_2}$ display
nonmonotonicities, which have drawn considerable theoretical
attention lately~\cite{Silvestrov00,Gefen04,Sindel05}. As seen
in Fig.~\ref{fig:LargeFeature}, the nonmonotonicities can be
quite large, although no population inversion occurs for
$\Gamma_{\uparrow} = \Gamma_{\downarrow}$.

Our general discussion in Sec.~\ref{sec:LocalMoment} makes it
is easy to interpret these features of the partial occupancies
$\qav{n_{i}}$. Indeed, as illustrated in
Fig.~\ref{fig:LargeFeature}, there is excellent agreement in
the local-moment regime between the exact Bethe \emph{ansatz}
results and the curves obtained from Eqs.~\eqref{eq:Actualn1n2}
and (\ref{eq:n0PT}) based on the mapping onto the Kondo
Hamiltonian. We therefore utilize Eqs.~\eqref{eq:Actualn1n2}
for analyzing the data. To begin with we note that, for
$\Gamma_{\spinup} = \Gamma_{\spindown}$, there is no
renormalization of the effective magnetic field. The latter
remains constant and equal to $h = \Delta$ independent of
$\epsilon_0$. Combined with the fact that $\cos(\theta_d +
s_\theta \theta_h) \equiv -1$ in Eqs.~\eqref{eq:Actualn1n2},
the magnetization $M = \qav{n_{\uparrow} - n_{\downarrow}}/2 =
\qav{n_{2} - n_{1}}/2$ depends exclusively on the ratio
$\Delta/T_K$. The sole dependence on $\epsilon_0$ enters
through $T_K$, which varies according to
Eq.~(\ref{eq:TKAndersonAccurate}). Thus, $M$ is positive for
all gate voltages $\epsilon_0$, excluding the possibility of a
population inversion.

The nonmonotonicities in the individual occupancies stem from
the explicit dependence of $T_K$ on the gate voltage
$\epsilon_0$. According to Eq.~(\ref{eq:TKAndersonAccurate}),
$T_K$ is minimal in the middle of the Coulomb-blockade valley,
increasing monotonically as a function of $|\epsilon_0 + U/2|$.
Thus, $\Delta/T_K$, and consequently $M$, is maximal for
$\epsilon_0 = -U/2$, decreasing monotonically the farther
$\epsilon_0$ departs from $-U/2$. Since $n_{\text{tot}} \approx
1$ is nearly a constant in the local-moment regime, this
implies the following evolution of the partial occupancies:
$\qav{n_1}$ decreases ($\qav{n_2}$ increases) as $\epsilon_0$
is lowered from roughly zero to $-U/2$. It then increases
(decreases) as $\epsilon_0$ is further lowered toward $-U$.
Combined with the crossovers to the empty-impurity and doubly
occupied regimes, this generates a local maximum (minimum) in
$\qav{n_1}$ ($\qav{n_2}$) near $\epsilon_0 \sim 0$ ($\epsilon_0
\sim -U$).

Note that the local extremum in $\qav{n_i}$ is most pronounced
for intermediate values of the field $\Delta$. This can be
understood by examining the two most relevant energy scales in
the problem, namely, the minimal Kondo temperature
$T_{K}^{\text{min}} = T_{K}^{}|_{\epsilon_0=-U/2}$ and the
hybridization width $\Gamma$. These two energies govern the
spin susceptibility of the impurity in the middle of the
Coulomb-blockade valley (when $\epsilon_0 = -U/2$) and in the
mixed-valent regime (when either $\epsilon_0 \approx 0$ or
$\epsilon \approx -U$), respectively. The charging curves of
Fig.~\ref{fig:LargeFeature} stem from an interplay of the three
energy scales $\Delta$, $T_K^{\text{min}}$ and $\Gamma$ as
described below.

\begin{figure}
\includegraphics[width=7.5cm]{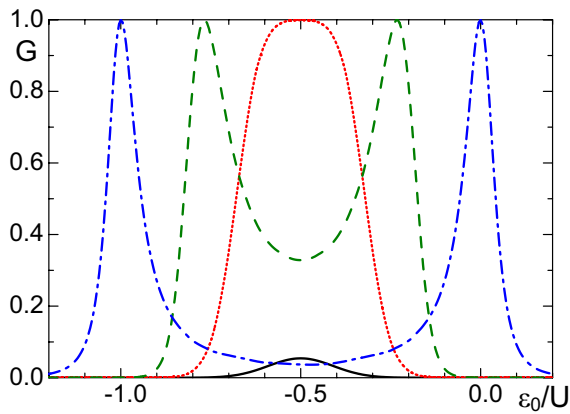}
\caption{(Color online) The exact conductance $G$
         [in units of $e^2/(2\pi\hbar)$] versus
         $\epsilon_0$, as obtained from the Bethe
         \emph{ansatz} magnetization $M$ and
         Eq.~(\ref{G-parallel}) with
         $\theta_l = 3\pi/2$ and $\theta_d = \pi$.
         Here $\Delta/U$ equals $10^{-5}$ [full
         (black) line], $10^{-4}$ [dotted (red)
         line], $10^{-3}$ [dashed (green) line]
         and $0.1$ [dot-dashed (blue) line]. The
         remaining model parameters are
         $\Gamma/U = 0.05$ and $T = 0$. Once $\Delta$
         exceeds the critical field $h_c^{} \approx
         2.4 T_K^{\text{min}}$, the single peak at
         $\epsilon_0 = -U/2$ is split into two
         correlation-induced peaks, which cross
         over to Coulomb-blockade peaks at large
         $\Delta$.}
\label{fig:isoCIR-1}
\end{figure}

When $\Delta \ll T_K^{\text{min}}$, exemplified by
the pair of curves corresponding to the smallest field
$\Delta = 10^{-5} U \approx 0.24 T^{\text{min}}_K$
in Fig.~\ref{fig:LargeFeature}, the magnetic field
remains small throughout the Coulomb-blockade valley
and no significant magnetization develops. The two
levels are roughly equally populated, showing a
plateaux at $\qav{n_{1}} \approx \qav{n_{2}}
\approx 1/2$ in the regime where the dot is singly
occupied. As $\Delta$ grows and approaches
$T_K^{\text{min}}$, the field becomes sufficiently strong
to significantly polarize the impurity in the vicinity
of $\epsilon_0 = -U/2$. A gap then rapidly develops
between $\qav{n_{1}}$ and $\qav{n_{2}}$ near
$\epsilon_0 = -U/2$ as $\Delta$ is increased. Once
$\Delta$ reaches the regime $T_K^{\text{min}} \ll
\Delta \ll \Gamma$, a crossover from $h \gg T_K$
(fully polarized impurity) to $h \ll T_K$ (unpolarized
impurity) occurs as $\epsilon_0$ is tuned away
from the middle of the Coulomb-blockade valley. This
leads to the development of a pronounced maximum
(minimum) in $\qav{n_1}$ ($\qav{n_2}$), as marked by
the arrows in Fig.~\ref{fig:LargeFeature}. Finally, when
$h \gtrsim \Gamma$, the field is sufficiently large
to keep the dot polarized throughout the local-moment
regime. The extremum in $\qav{n_i}$ degenerates into
a small bump in the vicinity of either $\epsilon_0
\approx 0$ or $\epsilon_0 \approx -U$, which is
the nonmonotonic feature first discussed in
Ref.~\onlinecite{Gefen04}. This regime is exemplified
by the pair of curves corresponding to the largest
field $\Delta = 0.1 U = 2\Gamma$ in
Fig.~\ref{fig:LargeFeature}, whose parameters match
those used in Fig.~2 of Ref.~\onlinecite{Gefen04}.
Note, however, that the perturbative calculations of
Ref.~\onlinecite{Gefen04} will inevitably miss the
regime $T_K^{\text{min}} \ll \Delta \ll \Gamma$ where
this feature is large~\cite{comm-Sindel-feature}.

\subsubsection{Conductance}
\label{Sec:isoCond}

\begin{figure}
\includegraphics[width=7cm]{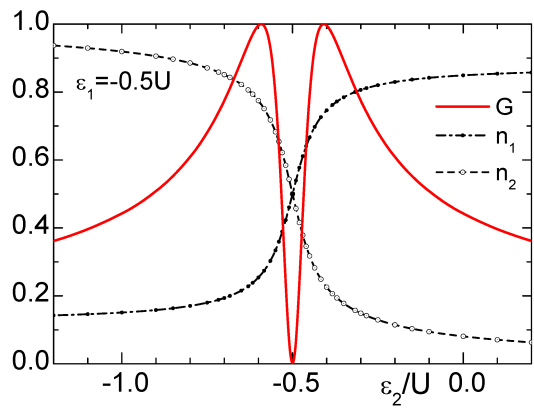}
\caption{(Color online) The exact occupation numbers
         $\qav{n_i}$ and conductance $G$ [in units
         of $e^2/(2 \pi \hbar)$] as a function of
         $\epsilon_2$, for $T = 0$, $\Gamma/U = 0.2$
         and fixed $\epsilon_1/U = -1/2$. The
         population inversion at $\epsilon_2 =
         \epsilon_1$ leads to a sharp transmission
         zero (phase lapse). Note the general
         resemblance between the functional dependence of
         $G$ on $\epsilon_2$ and the correlation-induced
         resonances reported by Meden and
         Marquardt~\cite{Meden06PRL} for
         $\Gamma_{\spinup} \neq \Gamma_{\spindown}$
         (see Fig.~\ref{fig:CIR}).}
\label{fig:isoCIR-2}
\end{figure}

The data of Fig.~\ref{fig:LargeFeature} can easily be
converted to conductance curves by using the exact
formula of Eq.~\eqref{G-parallel} with $\theta_l = 3\pi/2$
and $\theta_d = \pi$.
The outcome is presented in Fig.~\ref{fig:isoCIR-1}.
The evolution of $G(\epsilon_0)$ with increasing
$\Delta$ is quite dramatic. When $\Delta$ is small,
the conductance is likewise small with a shallow peak
at $\epsilon_0 = -U/2$. This peak steadily grows with
increasing $\Delta$ until reaching the unitary limit, at
which point it is split in two. Upon further increasing
$\Delta$, the two split peaks gradually depart,
approaching the peak positions $\epsilon_0 \approx 0$
and $\epsilon_0 \approx -U$ for large $\Delta$. The
conductance at each of the two maxima remains pinned
at all stages at the unitary limit.

These features of the conductance can be naturally
understood based on Eqs.~\eqref{G-parallel} and
\eqref{eq:Actualn1n2}. When $\Delta \ll T_K^{\text{min}}$,
the magnetization $M \approx \Delta/(2\pi T_K)$ and the
conductance $G \approx (\Delta/T_K)^2 e^2/(2\pi \hbar)$
are uniformly small, with a peak at $\epsilon_0 = -U/2$
where $T_K$ is the smallest. The conductance
monotonically grows with increasing $\Delta$ until
reaching the critical field $\Delta = h_c^{} \approx 2.4
T_K^{\text{min}}$, where $M|_{\epsilon_0 = -U/2} = 1/4$
and $G|_{\epsilon_0 = -U/2} = e^2/(2 \pi \hbar)$. Upon
further increasing $\Delta$, the magnetization at
$\epsilon_0 = -U/2$ exceeds $1/4$, and the associated
conductance decreases. The unitarity condition
$M = 1/4$ is satisfied at two gate voltages
$\epsilon^{\pm}_{\text{max}}$ symmetric about $-U/2$,
defined by the relation $T_K \approx \Delta/2.4$. From
Eq.~(\ref{eq:TKAndersonAccurate}) one obtains
\begin{equation}
\epsilon_{\pm}^{\text{max}} = -\frac{U}{2}
      \pm \sqrt{
                 \frac{U^2}{4} - \Gamma^2 +
                 \frac{2\Gamma U}{\pi}
                 \ln \left (
                             \frac{\pi \Delta}
                                  {2.4 \sqrt{2 \Gamma U}}
                     \right )
               } \, .
\end{equation}
The width of the two conductance peaks,
$\Delta \epsilon$, can be estimated for
$T_K^{\text{min}} \ll \Delta \ll \Gamma$ from the inverse
of the derivative $d(\Delta/T_K)/d\epsilon_0$, evaluated
at $\epsilon_0^{} = \epsilon_{\text{max}}^{\pm}$. It
yields
\begin{equation}
\Delta \epsilon \sim \frac{\Gamma U}
       {\pi | \epsilon_{\text{max}}^{\pm} + U/2 |} \, .
\end{equation}
Finally, when $\Delta > \Gamma$, the magnetization
exceeds $1/4$ throughout the local-moment regime. The
resonance condition $M = 1/4$ is met only as charge
fluctuations become strong, namely, for either
$\epsilon_0 \approx 0$ or $\epsilon_0 \approx -U$. The
resonance width $\Delta \epsilon$ evolves continuously in
this limit to the standard result for the Coulomb-blockade
resonances, $\Delta \epsilon \sim \Gamma$.

Up until now the energy difference $\Delta$ was kept
constant while tuning the average level position
$\epsilon_0$. This protocol, which precludes population
inversion as a function of the control parameter, best
suits a single-dot realization of our model, where both
levels can be uniformly tuned using a single gate voltage.
In the alternative realization of two spatially separated
quantum dots, each controlled by its own separate gate
voltage, one could fix the energy level
$\epsilon_1 = \epsilon_0 + \Delta/2$ and sweep the other
level, $\epsilon_2 = \epsilon_0 - \Delta/2$. This setup
amounts to changing the field $h$ externally, and is
thus well suited for probing the magnetic response of
our effective impurity.

An example for such a protocol is presented in
Fig.~\ref{fig:isoCIR-2}, where $\epsilon_1$ is held
fixed at $\epsilon_1 = -U/2$. As $\epsilon_2$ is
swept through $\epsilon_1$, a population inversion
takes place, leading to a narrow dip in the conductance.
The width of the conductance dip is exponentially
small due to Kondo correlations. Indeed, one can
estimate the dip width, $\Delta \epsilon_{\text{dip}}$,
from the condition $|\epsilon_1 - \epsilon_2| =
T_K|_{\epsilon_2 = \epsilon_1}$, which yields
\begin{equation}
\Delta \epsilon_{\text{dip}} \sim
       \sqrt{U \Gamma} \exp
                       \left (
                               -\frac{\pi U}{8 \Gamma}
                       \right ) \, .
\label{eq:CIRwidthIsotropic}
\end{equation}

\subsection{Anisotropic couplings, $\Gamma_{\uparrow}
            \neq \Gamma_{\downarrow}$}
\label{sec:ResultsAnisotropic}

As demonstrated at length in Sec.~\ref{sec:ResultsIsotropic},
the occurrence of population inversion and a transmission
zero for $\Gamma_{\uparrow} = \Gamma_{\downarrow}$
requires an external modulation of the effective
magnetic field. Any practical device will inevitably
involve, though, some tunnelling anisotropy,
$V_{\spinup} \neq V_{\spindown}$. The latter provides
a different route for changing the effective magnetic
field, through the anisotropy-induced terms in
Eq.~\eqref{h-total}. Implementing the same protocol
as in Sec.~\ref{sec:ResCharging} (that is, uniformly
sweeping the average level position $\epsilon_0$ while
keeping the difference $\Delta$ constant) would now
generically result both in population inversion and a
transmission zero due to the rapid change in direction
of the total field $\vec{h}_{\text{tot}}$. As emphasized
in Sec.~\ref{sec:occupany-PF}, the two phenomena
will generally occur at different gate voltages
when $V_{\spinup} \neq V_{\spindown}$.

\subsubsection{Degenerate levels, $\Delta = b = 0$}

We begin our discussion with the case where
$\Delta = b = 0$, which was extensively studied in
Ref.~\onlinecite{Meden06PRL}. It corresponds to a
particular limit of the parallel-field configuration where
$h = 0$. In the parallel-field configuration, the
conductance $G$ and occupancies $\qav{n_i}$ take the exact
forms specified in Eqs.~(\ref{G-parallel-field}) and
(\ref{eq:Actualn1n2-PF}), respectively. These expressions
reduce in the Kondo regime to Eqs.~(\ref{G-no-flux}) and
(\ref{eq:Actualn1n2}), with $\theta_h$ either equal to
zero or $\pi$, depending on the sign of $h_{\text{tot}}^z$.

\begin{figure}
\includegraphics[width=8cm]{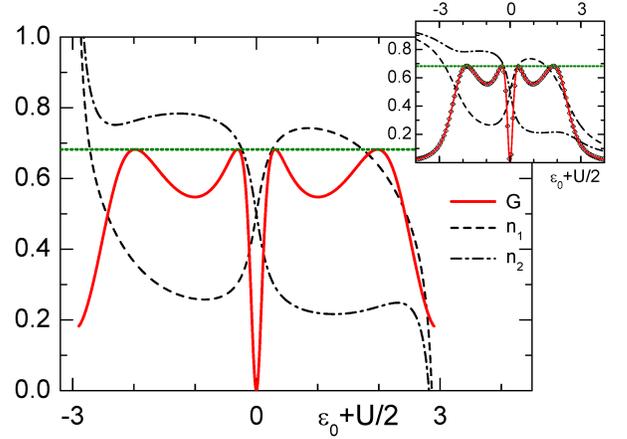}
\caption{The occupation numbers $\qav{n_i}$ and conductance
         $G$ [in units of $e^2/(2 \pi \hbar)$] as a
         function of $\epsilon_0 + U/2$ [in units of
         $\Gamma_{\text{tot}} = (\Gamma_{\spinup} +
         \Gamma_{\spindown})$],
         calculated from Eqs.~(\ref{G-no-flux}) and
         (\ref{eq:Actualn1n2}) based on the mapping
         onto the Kondo model. The model parameters
         are identical to those used in Fig.~2 of
         Ref.~\onlinecite{Meden06PRL}, lower left panel:
         $h = \varphi = 0$, $U/\Gamma_{\text{tot}} = 6$,
         $\Gamma_{\spinup}/\Gamma_{\text{tot}} = 0.62415$
         and $T = 0$. The
         explicit tunnelling matrix elements are detailed
         in Eq.~(\ref{eq:AMM}), corresponding to the
         rotation angles $\theta_l = 2.1698$ and
         $\theta_d = -0.63434$ (measured in radians).
         The angle $\theta_h$ equals zero. The
         inset shows functional renormalization-group (fRG)
         data as defined in Ref.~\onlinecite{Meden06PRL},
         corrected for the renormalization of the
         two-particle vertex~\cite{Karrasch06,MedenThanks}.
         The small symbols in the inset
         show the conductance as calculated
         from the fRG occupation numbers using our
         Eq.~\eqref{G-parallel}. The horizontal dotted
         lines in each plot mark the maximal conductance
         predicted by Eq.~\eqref{G-parallel},
         $(e^2/2 \pi \hbar) \sin^2 \theta_l$.}
\label{fig:CIR}
\end{figure}

Figure~\ref{fig:CIR} shows the occupation numbers and
the conductance obtained from Eqs.~(\ref{G-no-flux})
and (\ref{eq:Actualn1n2}), for $\Delta = b = 0$ and
the particular tunnelling matrix used in Fig.~2 of
Ref.~\onlinecite{Meden06PRL}:
\begin{align}
\hat{A} = A_0
          \begin{bmatrix}
                 \sqrt{0.27} & \sqrt{0.16} \\
                 \sqrt{0.33} & -\sqrt{0.24}\\
\end{bmatrix} \, .
\label{eq:AMM}
\end{align}
Here $A_0$ equals $\sqrt{\Gamma_{\text{tot}}/ (\pi \rho)}$,
with $\Gamma_{\text{tot}} =
\Gamma_{\spinup}+\Gamma_{\spindown}$ being the combined
hybridization width. The Coulomb repulsion $U$ is set equal to
$6 \Gamma_{\text{tot}}$, matching the value used in the lower
left panel of Fig.~2 in Ref.~\onlinecite{Meden06PRL}. For
comparison, the corresponding functional renormalization-group
(fRG) data of Ref.~\onlinecite{Meden06PRL} is shown in the
inset, after correcting for the renormalization of the
two-particle vertex~\cite{Karrasch06,MedenThanks}. The accuracy
of the fRG has been established~\cite{Meden06PRL,Karrasch06} up
to moderate values of $U/\Gamma_{\text{tot}} \sim 10$ through
a comparison with Wilson's numerical renormalization-group
method~\cite{NrgMethods}. Including the renormalization of
the two-particle vertex further improves the fRG data as
compared to that of Ref.~\onlinecite{Meden06PRL}, as reflected,
e.g., in the improved position of the outer pair of conductance
resonances.

The agreement between our analytical approach and the fRG is
evidently very good in the local-moment regime, despite the
rather moderate value of $U/\Gamma_{\text{tot}}$ used.
Noticeable deviations develop in $\qav{n_i}$ only as the
mixed-valent regime is approached (for $\epsilon_0 \agt
-\Gamma_{\text{tot}}$ or $\epsilon + U \alt
\Gamma_{\text{tot}}$), where our approximations naturally break
down. In particular, our approach accurately describes the
phase lapse at $\epsilon_0 = -U/2$, the inversion of population
at the same gate voltage, the location and height of the
correlation-induced resonances, and even the location and
height of the outer pair of conductance resonances. Most
importantly, our approach provides a coherent analytical
picture for the physics underlying these various features, as
elaborated below.

Before proceeding to elucidate the underlying physics, we
briefly quote the relevant parameters that appear in the
conversion to the generalized Anderson model of
Eq.~(\ref{eq:Hand}). Using the prescriptions detailed in
Appendix~\ref{App:SVDdetails}, the hybridization widths
$\Gamma^{}_{\sigma} = \pi \rho V_{\sigma}^2$ come out to be
\begin{equation}
\Gamma_{\uparrow}/\Gamma_{\text{tot}} = 0.62415 \; ,
\;\;\;\;
\Gamma_{\downarrow}/\Gamma_{\text{tot}} = 0.36585 \, ,
\end{equation}
while the angles of rotation equal
\begin{equation}
\theta_l = 2.1698 \; , \;\;\;\;
\theta_d = -0.63434 \, .
\end{equation}
Here $\theta_l$ and $\theta_d$ are quoted in radians. Using the
exact conductance formula of Eq.~(\ref{G-parallel-field}), $G$
is predicted to be bounded by the maximal conductance
\begin{equation}
G_{\text{max}} =
        \frac{e^2}{2 \pi \hbar} \sin^2 \theta_l
        = 0.68210 \frac{e^2}{2 \pi \hbar} \, ,
\label{G-max}
\end{equation}
obtained whenever the magnetization $M = \qav{n_{\uparrow} -
n_{\downarrow}}/2$ is equal to $\pm 1/4$. The heights of the
fRG resonances are in excellent agreement with
Eq.~(\ref{G-max}). Indeed, as demonstrated in the inset to
Fig.~\ref{fig:CIR}, the fRG occupancies and conductance comply
to within extreme precision with the exact relation of
Eq.~(\ref{G-parallel}). As for the functional form of the Kondo
temperature $T_K$, its exponential dependence on $\epsilon_0$
is very accurately described by Eq.~(\ref{scaling-T_K-2}). In
the absence of a precise expression for the pre-exponential
factor when $\Gamma_{\uparrow} \neq \Gamma_{\downarrow}$, we
employ the expression
\begin{equation}
T_K = (\sqrt{U \Gamma_{\text{tot}}}/\pi) \exp
  \left [
           \frac{\pi \epsilon_0 (U + \epsilon_0)}
                {2U(\Gamma_{\uparrow}-\Gamma_{\downarrow})}
           \ln\!
           \frac{\Gamma_{\uparrow}}{\Gamma_{\downarrow}}
  \right ] \, ,
\label{T_K-anisotropic}
\end{equation}
which properly reduces to Eq.~\eqref{eq:TKAndersonAccurate} (up
to the small $\Gamma^2$ correction in the exponent) when
$\Gamma_{\uparrow} = \Gamma_{\downarrow} = \Gamma$.

The occupancies and conductance of Fig.~\ref{fig:CIR} can be
fully understood from our general discussion in
Sec.~\ref{sec:LocalMoment}. Both quantities follow from the
magnetization $M$, which vanishes at $\epsilon_0 = -U/2$ due to
particle-hole symmetry. As a consequence, the two levels are
equally populated at $\epsilon_0 = -U/2$ and the conductance
vanishes [see Eqs.~(\ref{G-parallel-field}) and
(\ref{eq:Actualn1n2-PF})]. Thus, there is a simultaneous phase
lapse and an inversion of population at $\epsilon_0 = -U/2$,
which is a feature generic to $\Delta = b = 0$ and arbitrary
$\hat{A}$. As soon as the gate voltage is removed from $-U/2$,
i.e., $\epsilon_0 = -U/2 + \delta \epsilon$ with $\delta
\epsilon \neq 0$, a finite magnetization develops due to the
appearance of a finite effective magnetic field
$\vec{h}_{\text{tot}} = h^z_{\text{tot}} \hat{z}$ with
\begin{equation}
h^{z}_{\text{tot}} \approx
       \frac{\Gamma_{\uparrow} - \Gamma_{\downarrow}}{\pi}
       \ln \frac{1 + 2\delta \epsilon/U}
                          {1 - 2\delta \epsilon/U}
\label{h-z-tot}
\end{equation}
[see Eq.~(\ref{eq:htotExplicit})]. Note that the sign of
$h^{z}_{\text{tot}}$ coincides with that of $\delta \epsilon$,
hence $M$ is positive (negative) for $\epsilon_0 > -U/2$
($\epsilon_0 < -U/2$). Since $\cos \theta_d > 0$ for the model
parameters used in Fig.~\ref{fig:CIR}, it follows from
Eq.~(\ref{eq:Actualn1n2-PF}) that $\qav{n_1} > \qav{n_2}$
($\qav{n_1} < \qav{n_2}$) for $\epsilon_0 > -U/2$ ($\epsilon_0
< -U/2$), as is indeed found in Fig.~\ref{fig:CIR}. Once again,
this result is generic to $\Delta = b = 0$, except for the sign
of $\cos \theta_d$ which depends on details of the tunnelling
matrix $\hat{A}$.

In contrast with the individual occupancies, the conductance $G$
depends solely on the magnitude of $M$, and is therefore a
symmetric function of $\delta \epsilon$. Similar to the rich
structure found for $\Gamma_{\spinup} = \Gamma_{\spindown}$
and $\Delta > 0$ in Fig.~\ref{fig:isoCIR-1}, the intricate
conductance curve in Fig.~\ref{fig:CIR} is the result of the
interplay between $h_{\text{tot}}^z$ and $T_K$, and the
nonmonotonic dependence of $G$ on $|M|$. The basic physical
picture is identical to that in Fig.~\ref{fig:isoCIR-1}, except
for the fact that the effective magnetic field
$h_{\text{tot}}^z$ is now itself a function of the gate voltage
$\epsilon_0$.

As a rule, the magnetization $|M|$ first increases with
$|\delta \epsilon|$ due to the rapid increase in
$h_{\text{tot}}^z$. It reaches its maximal value
$M_{\text{max}}$ at some intermediate $|\delta \epsilon|$
before decreasing again as $|\delta \epsilon|$ is further
increased. Inevitably $|M|$ becomes small again once $|\delta
\epsilon|$ exceeds $U/2$. The shape of the associated
conductance curve depends crucially on the magnitude of
$M_{\text{max}}$, which monotonically increases as a function
of $U$. When $M_{\text{max}} < 1/4$, the conductance features
two symmetric maxima, one on each side of the particle-hole
symmetric point. Each of these peaks is analogous to the one
found in Fig.~\ref{fig:isoCIR-1} for $\Delta < h_c$. Their
height steadily grows with increasing $U$ until the unitarity
condition $M_{\text{max}} = 1/4$ is met. This latter condition
defines the critical repulsion $U_c$ found in
Ref.~\onlinecite{Meden06PRL}. For $U > U_c$, the maximal
magnetization $M_{\text{max}}$ exceeds one quarter. Hence the
unitarity condition $M = \pm 1/4$ is met at two pairs of gate
voltages, one pair of gate voltages on either side of the
particle-hole symmetric point $\epsilon_0 = -U/2$. Each of the
single resonances for $U < U_c$ is therefore split in two, with
the inner pair of peaks evolving into the correlation-induced
resonances of Ref.~\onlinecite{Meden06PRL}. The point of
maximal magnetization now shows up as a local minimum of the
conductance, similar to the point $\epsilon_0 = -U/2$ in
Fig.~\ref{fig:isoCIR-1} when $\Delta > h_c$.

For large $U \gg \Gamma_{\text{tot}}$, the magnetization $|M|$
grows rapidly as one departs from $\epsilon_0 = -U/2$, due to
the exponential smallness of the Kondo temperature
$T_K|_{\epsilon_0 = -U/2}$.  The dot remains polarized
throughout the local-moment regime, loosing its polarization
only as charge fluctuations become strong. In this limit the
inner pair of resonances lie exponentially close to $\epsilon_0
= -U/2$ (see below), while the outer pair of resonances
approach $|\delta \epsilon| \approx U/2$ (the regime of
the conventional Coulomb blockade).

The description of this regime can be made quantitative by
estimating the position $\pm \delta \epsilon_{\text{CIR}}$ of
the correlation-induced resonances. Since $M \to
M_K(h^{z}_{\text{tot}}/T^{}_K)$ deep in the local-moment
regime, and since $\delta \epsilon_{\text{CIR}} \ll
\Gamma_{\text{tot}}$ for $\Gamma_{\text{tot}} \ll U$, the
correlation-induced resonances are peaked at the two gate
voltages where $h^{z}_{\text{tot}} \approx \pm 2.4
T^{}_K|_{\epsilon_0 = -U/2}$. Expanding Eq.~(\ref{h-z-tot}) to
linear order in $\delta \epsilon_{\text{CIR}}/U \ll 1$ and
using Eq.~(\ref{T_K-anisotropic}) one finds
\begin{eqnarray}
\delta \epsilon_{\text{CIR}} &\approx& 0.6
       \frac{\pi U}{\Gamma_{\spinup}-\Gamma_{\spindown}}
       T_K|_{\epsilon_0 = -U/2}
\nonumber \\
&=& 0.6 \frac{U \sqrt{U \Gamma_{\text{tot}}}}
             {\Gamma_{\spinup}-\Gamma_{\spindown}}
      \exp\!
      \left [
              \frac{-\pi U \ln(\Gamma_{\spinup}/
                               \Gamma_{\spindown})}
                   {8(\Gamma_{\spinup}-\Gamma_{\spindown})}
      \right ] .
\label{eq:CIRwidth}
\end{eqnarray}
Here the pre-exponential factor in the final expression for
$\delta \epsilon_{\text{CIR}}$ is of the same accuracy as that
in Eq.~(\ref{T_K-anisotropic}).

We note in passing that the shape of the correlation-induced
resonances and the intervening dip can be conveniently
parameterized in terms of the peak position $\delta
\epsilon_{\text{CIR}}$ and the peak conductance
$G_{\text{max}}$. Expanding Eq.~(\ref{h-z-tot}) to linear order
in $\delta \epsilon/U \ll 1$ and using
Eq.~(\ref{G-parallel-field}) one obtains
\begin{equation}
G(\delta \epsilon) = G_{\text{max}} \sin^2\!
  \left [
          2 \pi M_K\!
            \left (
                     \frac{2.4 \delta \epsilon}
                          {\delta \epsilon_{\text{CIR}}}
            \right )
  \right ] \, ,
\end{equation}
where $M_K(h/T_K)$ is the universal magnetization curve of the
Kondo model [given explicitly by \eqref{eq:MKfullWiegmann}].
This parameterization in terms of two easily extractable
parameters may prove useful for analyzing future experiments.

It is instructive to compare Eq.~(\ref{eq:CIRwidth}) for
$\delta \epsilon_{\text{CIR}}$ with the fRG results of
Ref.~\onlinecite{Meden06PRL}, which tend to overestimate
$\delta \epsilon_{\text{CIR}}$. For the special case where
$a_{L 1} = a_{R 1}$ and $a_{L 2} = -a_{R 2}$, an analytic
expression was derived for $\delta \epsilon_{\text{CIR}}$ based
on the fRG~\cite{Meden06PRL}. The resulting expression,
detailed in Eq.~(4) of Ref.~\onlinecite{Meden06PRL}, shows an
exponential dependence nearly identical to that of
Eq.~(\ref{eq:CIRwidth}), but with an exponent that is smaller
in magnitude by a factor of $\pi^2/8 \approx
1.23$~\cite{Relating-the-Gamma's}. The same numerical factor
appears to distinguish the fRG and the numerical
renormalization-group data depicted in Fig.~3 of
Ref.~\onlinecite{Meden06PRL}, supporting the accuracy of our
Eq.~(\ref{eq:CIRwidth}). It should be emphasized, however, that
Fig.~3 of Ref.~\onlinecite{Meden06PRL} pertains to the
tunnelling matrix of Eq.~(\ref{eq:AMM}) rather than the special
case referred to above.

We conclude the discussion of the case where $\Delta = b = 0$
with accurate results on the renormalized dot levels when the
dot is tuned to the peaks of the correlation-induced
resonances. The renormalized dot levels,
$\tilde{\epsilon}_{\uparrow}$ and
$\tilde{\epsilon}_{\downarrow}$, can be defined through the $T
= 0$ retarded dot Green functions at the Fermi energy:
\begin{equation}
G_{\sigma}(\epsilon = 0) =
           \frac{1}{-\tilde{\epsilon}_{\sigma}
                      + i\Gamma_{\sigma}} \, .
\label{renormalized-levels-def}
\end{equation}
Here, in writing the Green functions of
Eq.~(\ref{renormalized-levels-def}), we have made use of the
fact that the imaginary parts of the retarded dot
self-energies, $-\Gamma_{\sigma}$, are unaffected by the
Coulomb repulsion $U$ at zero temperature at the Fermi energy.
The energies $\tilde{\epsilon}_{\sigma}$ have the exact
representation~\cite{Langreth66} $\tilde{\epsilon}_{\sigma} =
\Gamma_{\sigma} \cot \delta_{\sigma}$ in terms of the
associated phase shifts $\delta_{\sigma} = \pi \qav{n_{\sigma}}$.
Since $M = \pm 1/4$ at the peaks of the correlation-induced
resonances, this implies that $\delta_{\sigma} = \pi/2 \pm
\sigma \pi/4$, where we have set $n_{\text{tot}} =
1$~\cite{Comment-on-renormalized-levels}. Thus, the
renormalized dot levels take the form
$\tilde{\epsilon}_{\sigma} = \mp\sigma \Gamma_{\sigma}$,
resulting in
\begin{equation}
\tilde{\epsilon}_{\uparrow} \tilde{\epsilon}_{\downarrow}
      = -\Gamma_{\uparrow} \Gamma_{\downarrow} \, .
\label{renormalized-levels}
\end{equation}

The relation specified in Eq.~(\ref{renormalized-levels})
was found in Ref.~\onlinecite{Meden06PRL},
for the special case where $a_{L 1} = a_{R 1}$ and
$a_{L 2} = -a_{R 2}$~\cite{Relating-the-Gamma's}.
Here it is seen to be a generic feature of the
correlation-induced resonances for $\Delta = b = 0$
and arbitrary $\hat{A}$.

\subsubsection{Nondegenerate levels:
               arbitrary $\Delta$ and $b$}

Once $\sqrt{\Delta^2 + b^2} \neq 0$, the conductance and the
partial occupancies can have a rather elaborate dependence on
the gate voltage $\epsilon_0$. As implied by the general
discussion in Sec.~\ref{sec:LocalMoment}, the underlying
physics remains driven by the competing effects of the
polarizing field $h_{\text{tot}}$ and the Kondo temperature
$T_K$. However, the detailed dependencies on $\epsilon_0$ can be
quite involving and not as revealing. For this reason we shall
not seek a complete characterization of the conductance $G$ and
the partial occupancies $\qav{n_i}$ for arbitrary couplings.
Rather, we shall focus on the case where no Aharonov-Bohm
fluxes are present and ask two basic questions: (i) under what
circumstances is the phenomenon of a phase lapse generic? (ii)
under what circumstances is a population inversion generic?

When $\varphi_L = \varphi_R = 0$, the conductance and the
partial occupancies are given by Eqs.~(\ref{G-no-flux}) and
(\ref{eq:Actualn1n2}), respectively. Focusing on $G$ and on
$\qav{n_1 - n_2}$, these quantities share a common form, with
factorized contributions of the magnetization $M_K$ and the
rotation angles. The factors containing
$M_K(h_{\text{tot}}/T_K)$ never vanish when $h \sin \theta \neq
0$, since $h_{\text{tot}}$ always remains positive. This
distinguishes the generic case from the parallel-field
configuration considered above, where phase lapses and
population inversions are synonymous with $M = 0$. Instead, the
conditions for phase lapses and population inversions to occur
become distinct once $h \sin \theta \neq 0$, originating from
the independent factors where the rotation angles appear. For a
phase lapse to develop, the combined angle $\theta_l + s_R
\theta_h$ must equal an integer multiple of $\pi$. By contrast,
the inversion of population requires that $\theta_d +
s_{\theta}\theta_h = \pi/2 \!\!\mod\!\pi$. Here the dependence
on the gate voltage $\epsilon_0$ enters solely through the
angle $\theta_h$, which specifies the orientation of the
effective magnetic field $\vec{h}_{\text{tot}}$ [see
Eq.~(\ref{eq:htotExplicit})]. Since the rotation angles
$\theta_l$ and $\theta_d$ are generally unrelated, this implies
that the two phenomena will typically occur, if at all, at
different gate voltages.

For phase lapses and population inversions to be ubiquitous,
the angle $\theta_h$ must change considerably as $\epsilon_0$
is swept across the Coulomb-blockade valley. In other words,
the effective magnetic field $\vec{h}_{\text{tot}}$ must nearly
flip its orientation in going from $\epsilon_0 \approx 0$ to
$\epsilon_0 \approx -U$. Since the $x$ component of the field
is held fixed at $h_{\text{tot}}^{x} = h \sin \theta > 0$, this
means that its $z$ component must vary from $h_{\text{tot}}^{z}
\gg h_{\text{tot}}^{x}$ to $-h_{\text{tot}}^{z} \gg
h_{\text{tot}}^{x}$ as a function of $\epsilon_0$. When this
requirement is met, then both a phase lapse and an inversion of
population are essentially guaranteed to occur. Since
$h_{\text{tot}}^{z}$ crudely changes by
\begin{equation}
\Delta h_{\text{tot}}^{z} \sim
    \frac{2}{\pi} (\Gamma_{\uparrow} - \Gamma_{\downarrow})
    \ln ( U/\Gamma_{\text{tot}} )
\end{equation}
as $\epsilon_0$ is swept across the Coulomb-blockade
valley, this leaves us with the criterion
\begin{equation}
(\Gamma_{\uparrow} - \Gamma_{\downarrow})
         \ln ( U/\Gamma_{\text{tot}} ) \gg
         \sqrt{\Delta^2 + b^2} \, .
\label{condition-for-PL}
\end{equation}
Conversely, if $\sqrt{\Delta^2 + b^2}\gg (\Gamma_{\uparrow} -
\Gamma_{\downarrow}) \ln ( U/\Gamma_{\text{tot}} )$, then
neither a phase lapse nor an inversion of population will occur
unless parameters are fine tuned. Thus, the larger $U$ is, the
more ubiquitous phase lapses
become~\cite{Golosov06,Meden06PRL}.

Although the logarithm $\ln(U/\Gamma_{\text{tot}})$ can be made
quite large, in reality we expect it to be a moderate factor of
order one. Similarly, the difference in widths
$\Gamma_{\uparrow} - \Gamma_{\downarrow}$ is generally expected
to be of comparable magnitude to $\Gamma_{\uparrow}$. Under
these circumstances, the criterion specified in
Eq.~(\ref{condition-for-PL}) reduces to $\Gamma_{\uparrow} \gg
\sqrt{\Delta^2 + b^2}$. Namely, phase lapses and population
inversions are generic as long as the (maximal) tunnelling rate
exceeds the level spacing. This conclusion is in line with that
of a recent numerical study of multi-level quantum
dots~\cite{Karrasch06num}.

Finally, we address the effect of nonzero $h = \sqrt{\Delta^2 +
b^2}$ on the correlation-induced resonances. When $h \gg
\Gamma_{\uparrow} \ln(U/\Gamma_{\text{tot}})$, the effective
magnetic field $h_{\text{tot}} \approx h$ is large throughout
the local-moment regime, always exceeding
$\Gamma_{\uparrow}$ and $\Gamma_{\downarrow}$. Consequently,
the dot is nearly fully polarized for all $-U < \epsilon_0 <
0$, and the correlation-induced resonances are washed out.
Again, for practical values of $U/\Gamma_{\text{tot}}$ this
regime can equally be characterized by $h \gg
\Gamma_{\uparrow}$~\cite{Meden06PRL}.

The picture for $\Gamma_{\uparrow}\ln(U/\Gamma_{\text{tot}})
\gg h$ is far more elaborate. When $T_K|_{\epsilon_0 = -U/2}
\gg h$, the magnetic field is uniformly small, and no
significant modifications show up as compared with the case
where $h = 0$. This leaves us with the regime $T_K|_{\epsilon_0
= -U/2} \ll h \ll \Gamma_{\uparrow}$, where various behaviors
can occur. Rather than presenting an exhaustive discussion of
this limit, we settle with identifying certain generic features
that apply when both components $|h \cos \theta|$ and $h
\sin\theta$ exceed $T_K|_{\epsilon_0 = -U/2}$. To begin with,
whatever remnants of the correlation-induced resonances that
are left, these are shifted away from the middle of the
Coulomb-blockade valley in the direction where
$|h_{\text{tot}}^z|$ acquires its minimal value. Consequently,
$h_{\text{tot}}$ and $T_K$ no longer obtain their minimal
values at the same gate voltage $\epsilon_0$. This has the
effect of generating highly asymmetric structures in place of
the two symmetric resonances that are found for $h = 0$. The
heights of these features are governed by the ``geometric''
factors $\sin^2 (\theta_l + s_R \theta_h)$ at the corresponding
gate voltages. Their widths are controlled by the underlying
Kondo temperatures, which can differ substantially in
magnitude. Since the entire structure is shifted away from the
middle of the Coulomb-blockade valley where $T_K$ is minimal,
all features are substantially broadened as compared with the
correlation-induced resonances for $h = 0$. Indeed, similar
tendencies are seen in Fig.~5 of Ref.~\onlinecite{Meden06PRL},
even though the model parameters used in this figure lie on the
borderline between the mixed-valent and the local-moment
regimes.

\section{Concluding remarks}
\label{sec:conclusions}

We have presented a comprehensive investigation of the general
two-level model for quantum-dot devices. A proper choice of the
quantum-mechanical representation of the dot and the lead
degrees of freedom reveals an exact mapping onto a generalized
Anderson model. In the local-moment regime, the latter
Hamiltonian is reduced to an anisotropic Kondo model with a
tilted effective magnetic field. As the anisotropic Kondo model
flows to the isotropic strong-coupling fixed point, this
enables a unified description of all coupling regimes of the
original model in terms of the universal magnetization curve of
the conventional isotropic Kondo model, for which exact results
are available. Various phenomena, such as phase lapses in the
transmission phase,~\cite{Silva02,Golosov06} charge
oscillations,~\cite{Gefen04,Sindel05} and correlation-induced
resonances~\cite{Meden06PRL,Karrasch06} in the conductance, can
thus be accurately and coherently described within a single
physical framework.

The enormous reduction in the number of parameters in
the system was made possible by the key observation that
a general, possibly non-Hermitian tunnelling matrix
$\hat{A}$ can always be diagonalized with the help of
two simultaneous unitary transformations, one pertaining
the dot degrees of freedom, and the other applied to
the lead electrons. This transformation, known as the
singular-value decomposition, should have
applications in other physical problems involving
tunnelling or transfer matrices without any special
underlying symmetries.

As the two-level model for transport is quite general, it can
potentially be realized in many different ways. As already
noted in the main text, the model can be used to describe
either a single two-level quantum dot or a double quantum dot
where each dot harbors only a single level. Such realizations
require that the spin degeneracy of the electrons will be
lifted by an external magnetic field. Alternative realizations
may directly involve the electron spin. For example, consider a
single spinful level coupled to two ferromagnetic leads with
\emph{non-collinear} magnetizations. Written in a basis with a
particular \emph{ad hoc} local spin quantization axis, the
Hamiltonian of such a system takes the general form of
Eq.~\eqref{IHAM}, after properly combining the electronic
degrees of freedom in both leads. As is evident from our
discussion, the local spin will therefore experience an
effective magnetic field that is not aligned with either of the
two magnetizations of the leads. This should be contrasted with
the simpler configurations of parallel and antiparallel
magnetizations, as considered, e.g., in
Refs.~\onlinecite{Martinek03PRL,MartinekNRGferro}
and~~\onlinecite{MartinekPRB05}.

Another appealing system for the experimental observation of the
subtle correlation effects discussed in the present paper is a
carbon nanotube-based quantum dot. In such a device both charging
energy and single-particle level spacing can be
sufficiently large~\cite{Buitelaar02} to provide a set of
well-separated discrete electron states. Applying external
magnetic field either perpendicular~\cite{Nygard00} or/and
parallel~\cite{HerreroSU4} to the nanotube gives great
flexibility in tuning the energy level structure, and thus
turns the system into a valuable testground for probing
the Kondo physics addressed in this study.

Throughout this paper we confined ourselves to spinless
electrons, assuming that spin degeneracy has been lifted by an
external magnetic field. Our mapping can equally be applied to
spinful electrons by implementing an identical singular-value
decomposition to each of the two spin orientations separately
(assuming the tunnelling term is diagonal in and independent of
the spin orientation). Indeed, there has been considerable
interest lately in spinful variants of the Hamiltonian of
Eq.~(\ref{IHAM}), whether in connection with lateral quantum
dots,~\cite{Kondo2stage,HofstetterZarand04} capacitively
coupled quantum dots,~\cite{KondoSU4,LeHuretal05,Galpinetal06}
or carbon nanotube devices~\cite{AguadoPRL05}. Among the
various phenomena that have been discussed in these contexts,
let us mention SU(4) variants of the Kondo
effect~\cite{KondoSU4,LeHuretal05,AguadoPRL05}, and
singlet-triplet transitions with two-stage screening on the
triplet side~\cite{Kondo2stage,HofstetterZarand04}.

Some of the effects that have been predicted for the
spinful case were indeed observed in lateral semiconductor
quantum dots~\cite{vanderWiel02,Granger05} and in carbon nanotube
quantum dots~\cite{HerreroSU4}. Still, there remains a
distinct gap between the idealized models that have been
employed, in which simplified symmetries are often imposed
on the tunnelling term, and the actual experimental systems
that obviously lack these symmetries. Our mapping should
provide a much needed bridge between the idealized models
and the actual experimental systems. Similar to the present
study, one may expect a single unified description
encompassing all coupling regimes in terms of just
a few basic low-energy scales. This may provide valuable
guidance for analyzing future experiments on such
devices.

\begin{acknowledgments}
The authors are thankful to V. Meden for kindly providing the
numerical data for the inset in Fig.~\ref{fig:CIR}. VK is
grateful to Z.~A.~N\'{e}meth for stimulating discussions of
perturbative calculations. This research was supported by a
Center of Excellence of the Israel Science Foundation, and by a
grant from the German Federal Ministry of Education and
Research (BMBF) within the framework of the German-Israeli
Project Cooperation (DIP). We have recently become aware of a
related study by Silvestrov and Imry,~\cite{SI-06} which
independently develops some of the ideas presented in this
work.
\end{acknowledgments}

\appendix
\section{Mapping parameters}
\label{App:SVDdetails}

In this Appendix we give the details of the mapping of
the original Hamiltonian, Eq.~(\ref{IHAM}), onto the
generalized Anderson Hamiltonian of Eq.~(\ref{eq:Hand}).

The first step is the diagonalization of the matrix
$\hat{A}$, Eq.~(\ref{AA}), which describes the coupling
between the dot and the leads in the original model. Since
$\hat{A}$ is generally complex and of no particular
symmetry, it cannot be diagonalized by a single similarity
transformation. Rather, two (generally different) unitary
matrices, $R_d$ and $R_l$, are required to achieve a
diagonal form,
\begin{equation}
    \left [
            \begin{array}{cc}
                  V_{\spinup} & 0 \\
                  0 & V_{\spindown}
            \end{array}
    \right ] =
    R^{}_l \, \hat{A} \, R_d^{\dagger} \ .
\label{eq:Vdef}
\end{equation}
This representation, known as the singular-value
decomposition, is a standard routine in numerical
packages. Here we provide a fully analytical treatment
of the $2\times2$ case relevant to our discussion. To
this end we parametrize the two rotation matrices
in the form
\begin{eqnarray}
R_{d} &=&
      e^{i\chi_A} e^{i (\chi_d/2) \tau_z}
      U(\theta_d, \phi_d) \, ,
\label{eq:phased} \\
R_{l} &=&
      e^{i (\chi_l/2) \tau_z} \,
      U(\theta_l, \phi_l) \, ,
\label{eq:phasel}
\end{eqnarray}
where
\begin{eqnarray}
U(\theta, \phi) \equiv
    \left [
       \begin{array}{cc}
          \cos(\theta/2) & e^{-i\phi} \sin(\theta/2) \\
          -e^{i\phi} \sin(\theta/2) & \cos(\theta/2)
       \end{array}
    \right ]
\label{eq:Ugeneral}
\end{eqnarray}
describes a rotation by angle $\theta$ about the axis
$-\sin(\phi)\,\hat{x} + \cos(\phi)\,\hat{y}$.

\begin{figure}
\includegraphics[width=6cm]{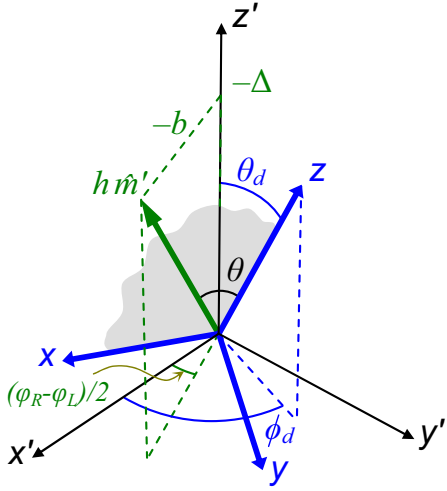}
\caption{The original dot degrees of freedom,
         $d^{\dagger}_1$ and $d^{\dagger}_2$, define a
         pseudo-spin-$\frac{1}{2}$ representation with
         the axes $x'$, $y'$, and $z'$. The level
         indices $1$ and $2$ are identified in this
         representation with $\pm \frac{1}{2}$ spin
         projections on the $z'$ axis. The energy
         splitting $\Delta$ and the hopping $b$ combine
         to define the magnetic-field vector $h \hat{m}'$.
         The unitary matrix $R_{d}$ takes the spin to
         a new coordinate system whose axes are labelled
         with $x$, $y$, and $z$. The new ``spin'' labels
         $\sigma = \uparrow$ and $\sigma = \downarrow$
         represent $\pm \frac{1}{2}$ spin projections
         on the new $z$ axis, whose direction
         is defined by the longitudinal and
         the azimuthal angles $\theta_d$ and $\phi_d$.
         The new $x$ axis lies in the plane of vectors
         $\hat{z}$ and $h \hat{m}'$. A similar picture
         applies to the conduction-electron degrees of
         freedom, where the lead index $\alpha = L, R$
         plays the same role as the original level
         index $i = 1, 2$.
  \label{fig:VectorRotation}
  }
\end{figure}

The various parameters that enter Eqs.~(\ref{eq:phased})
and (\ref{eq:phasel}) have simple geometrical
interpretations. The two sets of angles,
$(\theta_d, \phi_d)$ and $(\theta_l, \phi_{l})$, are
the longitudinal and the azimuthal angles of the vectors
pointing along the direction of the $z$ axis which defines
the corresponding spin variables in Eq.~(\ref{eq:Hand}),
see Fig.~\ref{fig:VectorRotation}. The three angles
$\chi_A$, $\chi_{d}$, and $\chi_{l}$ correspond to the
choice of the phases of the single-particle operators
$d^{\dagger}_{\sigma}$ and $c_{k \sigma}^{\dagger}$. The
latter angles are chosen such that the matrix elements
of the transformed Hamiltonian, Eq.~(\ref{eq:Hand}),
will be real with $h \sin \theta \ge 0$. Note that
$R_{d}$ and $R_{l}$ are determined up to a common overall
phase. This degree of freedom has been exhausted in
Eqs.~(\ref{eq:phased}) and (\ref{eq:phasel}) by requiring
that $\det R_{l} = 1$.

In order to determine the rotation matrices $R_d$
and $R_l$, one diagonalizes the hermitian matrices
$\hat{A}\hat{A}^{\dagger}$ and $\hat{A}^{\dagger}
\hat{A}$, whose eigenvalues are evidently real
and equal to $|V_{\sigma}|^{2}$. This calculation
determines the matrices $U(\theta_d ,\phi_d)$ and
$U(\theta_{l},\phi_{l})$, and yields the values of
$|V_{\sigma}|$. Indeed, using Eqs.~(\ref{eq:Vdef}),
(\ref{eq:phased}), and (\ref{eq:phasel}), one obtains
\begin{eqnarray}
\left [
        \begin{array}{cc}
              |V_{\spinup}|^2 & 0 \\
              0 & |V_{\spindown}|^2
        \end{array}
\right ] &=&
        U(\theta_l, \phi_l) \,
        \hat{A} \hat{A}^{\dagger}
        \, U^{\dagger}(\theta_l, \phi_l)
\label{eq:lddiag}
\nonumber\\
         &=& U(\theta_d, \phi_d) \,
             \hat{A}^{\dagger} \hat{A} \,
             U^{\dagger}(\theta_d, \phi_d) .
\end{eqnarray}
Assuming $|V_{\uparrow}| > |V_{\downarrow}|$ (the case
where $|V_{\uparrow}| = |V_{\downarrow}|$ is treated
separately in Sec.~\ref{app:Isotropic-couplings}),
these two equations give
\begin{equation}
|V_{\sigma}|^2 = X \pm Y \, ,
\label{eq:eigenV}
\end{equation}
\begin{equation}
\theta_{d/l} = 2 \arctan
       \sqrt{\frac{Y - Z_{d/l}}{Y + Z_{d/l}}\ }
       \, ,
\label{eq:thetadef}
\end{equation}
\begin{equation}
\phi_{d} = \arctan
       \left (
               \frac{a_{L1} a_{L2} - a_{R1} a_{R2}}
                    {a_{L1} a_{L2} + a_{R1} a_{R2}}
               \tan \frac{\varphi}{2}
       \right) + \pi \, \eta_d \, ,
\label{eq:phi_d}
\end{equation}
\begin{equation}
\phi_{l} = \arctan
       \left (
               \frac{a_{L2} a_{R2} - a_{L1} a_{R1}}
                    {a_{L2} a_{R2} + a_{L1} a_{R1}}
                    \tan \frac{\varphi}{2}
       \right) + \pi \, \eta_l \, ,
\label{eq:phi_l}
\end{equation}
where
\begin{eqnarray}
X &=& \frac{1}{2} \sum_{\alpha i} a_{\alpha i }^2 \, ,
\label{eq:Xdef} \\
Y &=& \sqrt{X^2 - |\det \hat{A}|^2} \, ,
\label{eq:Ydef} \\
Z_d &=& \frac{1}{2} \sum_{\alpha = L, R}
        (a_{\alpha 1}^2-a_{\alpha 2}^2) \, ,
\label{Z-d-def} \\
Z_l &=& \frac{1}{2} \sum_{i = 1, 2}
        (a_{L i}^2 - a_{R i}^2) \, ,
\label{Z-l-def}
\end{eqnarray}
and
\begin{eqnarray}
2 \, \eta_d &=& 1 - \sgn
           \left [
                \left (
                        a_{L1} a_{L2} + a_{R1} a_{R2}
                \right ) \cos \frac{\varphi}{2}
           \right ] \, ,
\label{sign-d} \\
2 \, \eta_l &=& 1 - \sgn
           \left [
                \left (
                        a_{L1} a_{R1} + a_{L2} a_{R2}
                \right ) \cos \frac{\varphi}{2}
           \right ] \, .
\label{sign-l}
\end{eqnarray}
The plus sign in Eq.~\eqref{eq:eigenV} corresponds to
$V_{\spinup}$, since the spin-up direction is defined
as the one with the larger coupling, $|V_{\spinup}|^2
> |V_{\spindown}|^2$. The longitudinal angles
$0 \le \theta_{d}, \theta_{l} \le \pi$ are uniquely
defined by Eq.~\eqref{eq:thetadef}, while the
quadrants for the azimuthal angles $-\pi/2  < \phi_{d},
\phi_{l} \le 3 \pi/2$ must be chosen according to
Eqs.~(\ref{sign-d}) and (\ref{sign-l}). The auxiliary
quantities in Eqs.~\eqref{eq:Xdef}--\eqref{Z-l-def} obey
the inequalities $X \ge Y$ and $Y \ge |Z_{d/l}|$.

The next step is to determine the angles $\chi_A$,
$\chi_d$, and $\chi_l$ which come to assure, among
other things, that $V_{\uparrow} > V_{\downarrow}$
are both real and non-negative. Let us begin with
$\chi_A$. When $\det \hat{A} \neq 0$, i.e., for
$V_{\downarrow} > 0$, the angle $\chi_A$ is
uniquely determined by taking the determinants of
both sides of Eq.~(\ref{eq:Vdef}) and equating
their arguments. This yields
\begin{equation}
\chi_A = \frac{1}{2} \arg
                 \det \hat{A} \, .
\label{eq:chiA}
\end{equation}
When $\det \hat{A} = 0$, the angle $\chi_A$ can take
arbitrary values. This stems from the fact that
$V_{\downarrow} = 0$, and therefore $c_{k \downarrow}$
can be attached an arbitrary phase without affecting
the form of Eq.~(\ref{eq:Hand}). In this case we
choose $\chi_A = 0$.

Next we rotate the Hamiltonian term $\hat{\mathcal{E}}_d$,
which is the first term of the isolated dot Hamiltonian,
Eq.~(\ref{HDOT}). Upon converting to the rotated
dot operators $d^{\dagger}_{\uparrow}$ and
$d^{\dagger}_{\downarrow}$, the single-particle
term $\hat{\mathcal{E}}_d$ transforms according to
\begin{equation}
\hat{\mathcal{E}}_d \to R^{}_d \, \hat{\mathcal{E}}^{}_d \,
                        R^{\dagger}_d \ .
\end{equation}
Consider first the partial rotation $U(\theta_d, \phi_d)
\hat{\mathcal{E}}_d U^{\dagger}(\theta_d, \phi_d)$ [see
Eq.~(\ref{eq:phased})]. Writing
$\hat{\mathcal{E}}_d$ [as defined in Eq.~\eqref{HPS}]
in the form
\begin{equation}
\hat{\mathcal{E}}_d = \epsilon_0
          - \frac{h}{2} \, \hat{m}' \cdot \vec{\tau}
\end{equation}
with
\begin{equation}
h = \sqrt{ \Delta^2 + b^2 }
\end{equation}
and
\begin{equation}
\hat{m}' = -\frac{b}{h}
          \cos \frac{\varphi_L-\varphi_R}{2}
          \hat{x}
          + \frac{b}{h}
          \sin \frac{\varphi_L-\varphi_R}{2}
          \hat{y}
          - \frac{\Delta}{h} \hat{z} \ ,
\label{eq:m-hat}
\end{equation}
the partial rotation $U(\theta_d, \phi_d) \hat{\mathcal{E}}_d
U^{\dagger}(\theta_d, \phi_d)$ gives
\begin{equation}
\epsilon_0 - \frac{h}{2} \, \hat{m} \cdot \vec{\tau} \ ,
\end{equation}
where $\hat{m}$ is the unit vector obtained by
rotating $\hat{m}'$ by an angle $-\theta_d$ about
the axis $-\sin(\phi_d)\,\hat{x} +
\cos(\phi_d)\,\hat{y}$. Defining the angle
$\theta \in [0, \pi]$ which appears in Eq.~(\ref{eq:Hand})
according to $\cos \theta = m_{z}$, it follows
from simple geometry that
\begin{eqnarray}
\cos \theta = &-&\!\frac{\Delta}{h} \cos \theta_d
\nonumber\\
&-&\! \frac{b}{h} \sin \theta_d
    \cos [ \phi_d +(\varphi_L-\varphi_R)/2 ] \ .
\end{eqnarray}

The full transformation $R^{}_d \hat{\mathcal{E}}_d R^{\dagger}_d$
corresponds to yet another rotation of $\hat{m}$ by an
angle $-\chi_d$ about the $z$ axis. The angle $\chi_d$
is chosen such that the projection of $\hat{m}$ onto the
$xy$ plane is brought to coincide with the $x$ direction.
This fixes $\chi_d$ uniquely, unless $h \sin \theta$
happens to be zero (whether because $h = 0$ or because
$\theta$ is an integer multiple of $\pi$). When
$h \sin \theta = 0$, the angle $\chi_d$ can take
arbitrary values. Physically this stems from the fact
that spin-up and spin-down degrees of freedom can be
gauged separately within Eq.~(\ref{eq:Hand}). We choose
$\chi_d = 0$ in this case. The explicit expression
for $\chi_d$ when $h \sin \theta \neq 0$ is quite
cumbersome and will not be specified. As for the
remaining angle $\chi_l$, it is fixed by the requirement
that $V_{\sigma}$ will be real and non-negative.

Note that the conditions for the two exactly
solvable cases quoted in the
main text, Eqs.~(\ref{exact-a}) and (\ref{exact-b}),
are readily derived from our expressions for the
eigenvalues $V_{\sigma}$. The first case,
Eq.~(\ref{exact-a}), corresponds to $V_{\downarrow} = 0$,
which requires $\det \hat{A} = e^{i \varphi} a_{L1}
a_{R2} - a_{L2} a_{R1}=0$. This immediately leads
to Eq.~\eqref{exact-a}. The second solvable case,
Eq.~\eqref{exact-b}, corresponds to equal eigenvalues,
which implies $Y = 0$ [Eqs.~(\ref{eq:eigenV}),
(\ref{eq:Xdef}) and (\ref{eq:Ydef}) remain intact
for $|V_{\uparrow}| = |V_{\downarrow}|$].
By virtue of the inequalities
$Y \ge |Z_{d/l}|$, this necessitates that $Z_d$ and
$Z_l$ are both zero, which gives rise to the first two
conditions in Eq.~\eqref{exact-b}. The remaining
condition on the Aharonov-Bohm phase $\varphi$ follows
from substituting the first two conditions into the
definition of $Y$ and equating $Y$ to zero.

\subsection{No Aharonov-Bohm fluxes}
\label{app:NoABflux}

Of particular interest is the case where no Aharonov-Bohm
fluxes are present, $\varphi_L = \varphi_R = 0$. In the
absence of a real magnetic field that penetrates the
structure, the parameters that appear in the Hamiltonian
of Eq.~(\ref{IHAM}) are all real. This greatly simplifies
the resulting expressions for the rotation matrices $R_d$
and $R_l$, as well as for the model parameters that
appear in Eq.~(\ref{eq:Hand}).
In this subsection, we provide explicit expression for
these quantities in the absence of Aharonov-Bohm fluxes,
focusing on the case where $V_{\uparrow} > V_{\downarrow}$.
The case where $V_{\uparrow} = V_{\downarrow}$ is
treated separately in Sec.~\ref{app:Isotropic-couplings}.

As is evident from Eqs.~(\ref{eq:phi_d}) and (\ref{eq:phi_l}),
each of the azimuthal angles $\phi_d$ and $\phi_l$ is
either equal to $0$ or $\pi$ when $\varphi = 0$. (The
corresponding $y'$ and $y$ axes are parallel in
Fig.~\ref{fig:VectorRotation}.) It is
therefore advantageous to set both azimuthal angles
to zero at the expense of extending the range for the
longitudinal angles $\theta_{d}$ and $\theta_{l}$ from
$[0, \pi]$ to $(-\pi, \pi]$. Within this convention,
Eq.~(\ref{eq:thetadef}) is replaced with
\begin{equation}
\theta_{d/l} = 2 \, s_{d/l} \arctan
               \sqrt{\frac{Y - Z_{d/l}}{Y + Z_{d/l}}\ }
       \, ,
\label{eq:theta-no-AB}
\end{equation}
where
\begin{eqnarray}
s_{d} &=& \sgn \left (
                        a_{L1} a_{L2} + a_{R1} a_{R2}
               \right ) \, ,
\\
s_{l} &=& \sgn \left (
                        a_{L1} a_{R1} + a_{L2} a_{R2}
               \right ) \, .
\end{eqnarray}
Similarly, the unit vector $\hat{m}'$ of
Eq.~(\ref{eq:m-hat}) reduces to
\begin{equation}
\hat{m}' = -\frac{b}{h} \hat{x}
          - \frac{\Delta}{h} \hat{z} \ ,
\end{equation}
which results in
\begin{eqnarray}
\hat{m} &=& \left [
                      -\frac{b}{h} \cos \theta_d
                      + \frac{\Delta}{h} \sin \theta_d
              \right ] \hat{x}
\nonumber \\
         &-& \left [
                      \frac{b}{h} \sin \theta_d
                      + \frac{\Delta}{h} \cos \theta_d
              \right ] \hat{z}
\end{eqnarray}
and
\begin{equation}
\theta = \pi - \arccos
         \left (
                   \frac{b}{h} \sin \theta_d
                   + \frac{\Delta}{h} \cos \theta_d
         \right ) .
\end{equation}

Since the rotated unit vector $\hat{m}$ has no $y$
component, the angle $\chi_d$ is either equal to $0$
or $\pi$, depending on the sign of $m_x$. Assuming
$\det \hat{A} \neq 0$ and using Eqs.~\eqref{eq:Ugeneral}
and \eqref{eq:chiA}, one can write Eq.~\eqref{eq:phased}
in the form
\begin{equation}
R_d  = \bigl (\sgn \det \hat{A} \bigr)^{1/2} \,
        e^{i \pi (1-s_\theta) \tau_z/4} \,
        e^{i (\theta_d/2) \tau_y} \, ,
\label{R_d-no-AB}
\end{equation}
\begin{equation}
s_{\theta} = \sgn m_x = \sgn
             \left (
                     \Delta \sin\theta_d - b \cos\theta_d
             \right ) \, .
\label{eq:stheta}
\end{equation}
Note that the first exponent in Eq.~(\ref{R_d-no-AB})
is equal to $1$ for $s_{\theta}=+1$, and is equal
to $e^{i (\pi/2) \tau_z}$ for $s_{\theta} = -1$. If
$\det \hat{A} = 0$ we set $\sgn \det \hat{A} \to 1$
in Eq.~(\ref{R_d-no-AB}), while for
$\Delta \sin \theta_d = b \cos \theta_d$ we select
$s_{\theta} = +1$.

Proceeding to the remaining angle $\chi_l$, we note that $R_d$ of
Eq.~(\ref{R_d-no-AB}) is either purely real or purely imaginary,
depending on whether
\begin{equation}
s_R = s_{\theta} \, \sgn \det\hat{A}
\label{eq:sR}
\end{equation}
is positive or negative. Since both $e^{i (\theta_l/2)
\tau_y}$ and $\hat{A}$ are real matrices, then
$e^{i (\chi_l/2) \tau_z}$ must also be either purely
real or purely imaginary in tandem with $R_d$ in order
for the eigenvalues $V_{\uparrow}$ and $V_{\downarrow}$
to be real. This consideration dictates that $\chi_l$
is an integer multiple of $\pi$, with an even (odd)
integer for positive (negative) $s_R$. The end result
for $R_l$ is therefore
\begin{equation}
R_l = \eta_R \, e^{i \pi (1-s_R) \tau_z/4} \,
      e^{i (\theta_l/2) \tau_y} \, ,
\label{R_l-no-AB}
\end{equation}
Here $\eta_R = \pm 1$ is an overall phase which comes
to assure that the eigenvalues $V_{\uparrow}$ and
$V_{\downarrow}$ are non-negative.

\subsection{Isotropic couplings, $V_{\uparrow} =
            V_{\downarrow}$}
\label{app:Isotropic-couplings}

Our general construction of the rotation matrices $R_{d}$
and $R_{l}$ fails when $|V_{\uparrow}| = |V_{\downarrow}|
= V$. Equations~(\ref{eq:eigenV}), (\ref{eq:Xdef}) and
(\ref{eq:Ydef}) remain in tact for $|V_{\uparrow}| =
|V_{\downarrow}|$, however the angles $\theta_{d/l}$
and $\phi_{d/l}$ are ill-defined in
Eqs.~(\ref{eq:thetadef})--(\ref{eq:phi_l}). This reflects
the fact that the matrices $\hat{A}^{\dagger} \hat{A}$
and $\hat{A} \hat{A}^{\dagger}$ are both equal to $V^2$
times the unit matrix, hence any rotation matrix
$U(\theta, \phi)$ can be used to ``diagonalize'' them.
There are two alternatives for treating the isotropic
case where $|V_{\uparrow}| = |V_{\downarrow}|$. The
first possibility is to add an infinitesimal matrix
$\eta \hat{B}$ that lifts the degeneracy of
$|V_{\uparrow}|$ and $|V_{\downarrow}|$:
$\hat{A} \to \hat{A} + \eta \hat{B}$. Using the general
construction outlined above and implementing the limit
$\eta \to 0$, a proper pair of rotation matrices $R_{d}$
and $R_{l}$ are obtained. The other alternative is to
directly construct the rotation matrices $R_{d}$ and
$R_{l}$ pertaining to this case. Below we present this
second alternative.

A key observation for the isotropic case pertains to
the ``reduced'' matrix
\begin{equation}
\hat{T} =
\bigl( \det \hat{A} \bigr)^{-1/2}  \hat{A} \, ,
\end{equation}
which obeys
\begin{equation}
\hat{T}^{\dagger} \hat{T} = \hat{T}\hat{T}^{\dagger} = 1
      \; , \;\;\;\;\; \det \hat{T}  =1 \, .
\end{equation}
As a member
of the SU(2) group, $\hat{T}$ can be written in the form
\begin{equation}
\hat{T} = U(\theta_T, \phi_T) \,
          e^{i (\chi_T/2) \tau_z}
\end{equation}
with $\theta_T \in [0, \pi]$. Explicitly, the angles
$\theta_T$, $\phi_T$ and $\chi_T$ are given by
\begin{align}
\theta_T & = 2 \arccos
             \bigl(
                    |\det \hat{A}|^{-1/2} \, |a_{L 1}|
             \bigr ) \, , \\
\chi_T & = 2 \arg
           \bigl[
               ( \det \hat{A} )^{-1/2} \, a_{L 1}
            \bigr ] \, ,
\end{align}
and
\begin{equation}
\phi_T = \arg
           \bigl[
               ( \det \hat{A} )^{-1/2} \, a_{R 1}
           \bigr ]
           - \pi - \chi_T/2 \, .
\end{equation}
Exploiting the fact that $\det \hat{A} = V^2$, the
matrix $\hat{A}$ takes then the form
\begin{equation}
\hat{A} = V \, R^{\dagger}_{l} R^{}_{d} =
          V \,  e^{i \chi_A} \,
          U(\theta_T, \phi_T) \,
          e^{i (\chi_T/2) \tau_z} \, ,
\label{R_l-times-R_d}
\end{equation}
where the angle $\chi_A$ is defined in Eq.~(\ref{eq:chiA}).

Equation~(\ref{R_l-times-R_d}) determines the matrix
product $R_{l}^{\dagger} R_{d}^{}$. Any two rotation
matrices that satisfy the right-most equality in
Eq.~(\ref{R_l-times-R_d}) transform the tunnelling
matrix $\hat{A}$ to $V$ times the unit matrix, as is
required. The rotation matrix $R_d$ is subject to yet
another constraint, which stems from the requirement
that $h \sin \theta \ge 0$ in Eq.~(\ref{eq:Hand}). We
note that this constraint as well does not uniquely
determine the matrix $R_d$.~\cite{Comment-on-uniqueness}
Perhaps the simplest choice for $R_d$ is given by
\begin{equation}
R_d = e^{i \chi_A} e^{i (\chi_d/2) \tau_z}
\label{R_d-isotropic}
\end{equation}
with
\begin{equation}
\chi_d = \frac{1}{2}(\varphi_L - \varphi_R) +
         \frac{\pi}{2} (1 - \sgn b) \, ,
\end{equation}
which corresponds to
\begin{equation}
h \cos \theta = -\Delta \; , \;\;\;\;
h \sin \theta = |b| \, .
\end{equation}
Adopting the choice of Eq.~(\ref{R_d-isotropic}), the
rotation matrix $R_l$ takes the form
\begin{equation}
R_l = e^{i \tau_z (\chi_d - \chi_T)/2} \,
      U(\theta_T, -\phi_T) \, ,
\end{equation}
where $\theta_T$, $\phi_T$, $\chi_T$ and
$\chi_d$ are listed above.

\section{Bethe \emph{ansatz} formulae}
\label{app:Bethe}

In this appendix we gather for convenience all relevant details
of the exact Bethe \emph{ansatz} solutions for the impurity
magnetization in the isotropic Kondo and Anderson models in the
presence of a finite magnetic field. Extensive reviews of these
solutions (including the anisotropic Kondo model) are
available in the literature.~\cite{WiegmannC83,
WiegmannA83,AndreiRMP83} Here we only summarize the main
results of relevance to our analysis, and briefly comment on
the numerical procedure. We confine ourselves to zero
temperature, although explicit equations do exist also at
finite temperature. Throughout the Appendix we employ units in
which $\mu_B g = 1$.

\subsection{Isotropic Kondo model}

We begin the presentation with the case of a Kondo impurity,
before turning to the more elaborate case of an Anderson
impurity. As a function of the magnetic field $h$, the
magnetization of an isotropic spin-$\frac{1}{2}$ Kondo impurity
is given by the explicit expression [see, e.g., Eq.~(6.23) of
Ref.~\onlinecite{WiegmannC83}]
\begin{multline}
M(h) =
\\
\frac{-i}{4 \sqrt{\pi} }
       \int_{-\infty}^{+\infty} \!\!\!\! d \omega
       \frac{(i \omega + 0)^{i \omega/2 \pi} \,
             \sech \left (
                          \frac{\omega}{2}
                   \right )} {(\omega - i0)
             \Gamma(\frac{1}{2} + i \frac{\omega}{2 \pi})}
       \left (
            \frac{h}{2 \pi T_K}
       \right )^{i \omega /\pi} \!\!\!\! .
\label{eq:MKfullWiegmann}
\end{multline}
Here $\Gamma(z)$ is the complex gamma function. The Kondo
temperature, $T_K$, is defined via the inverse of the spin
susceptibility,
\begin{equation}
T_K^{-1} \equiv 2 \pi \lim_{h\to0} M(h)/h \, .
\label{eq:TKdef}
\end{equation}
Evidently, Eq.~(\ref{eq:MKfullWiegmann}) is a universal
function of the ratio $h/T_K$, which is denoted in the main
text by $M_K(h/T_K)$. It has the asymptotic expansion
\begin{align}
M(h) \simeq
     \begin{cases}
           h/(2\pi T_K) \, ,
           & h \ll T_K \, , \\
           \frac{1}{2} - \frac{1}{4 \ln (h/T_H)} -
           \frac{\ln \ln (h/T_H)}{8 \ln^2 (h/T_H)} \, ,
           & h \gg T_K \, , \\
     \end{cases}
\label{eq:MKassympt}
\end{align}
where $T_H \equiv \sqrt{\pi/e} T_K$.

\subsection{Isotropic Anderson model}

In contrast to the Kondo model, there are no closed-form
expressions for the total impurity occupancy $n_{\text{tot}} =
\qav{n_{\spinup} + n_{\spindown}}$ and magnetization $M =
\qav{n_{\spinup} - n_{\spindown}}/2$ in the isotropic Anderson
model. The exact Bethe \emph{anstaz} solution of the model
provides a set of coupled linear integral equations from which
$n_{\text{tot}}$ and $M$ can be computed. Below we summarize
the equations involved and comment on the numerical procedure
that is required for solving these equations. The expressions
detailed below apply to arbitrary $\epsilon_0$, $U$, $h$, and
$\Gamma$ at zero temperature. In terms of the Hamiltonian of
Eq.~(\ref{eq:Hand}), we restrict the discussion to $\Gamma =
\Gamma_{\uparrow} = \Gamma_{\downarrow}$ (isotropic Anderson
model) and $\sin \theta = 0$ (parallel-field configuration).
The case $\sin \theta \neq 0$ follows straightforwardly from a
simple rotation of the dot and the conduction-electron
operators about the $y$ axis.

\subsubsection{Kondo temperature}

The most accurate analytical expression that is available for
the Kondo temperature of the isotropic Anderson model can be
written as
\begin{equation}
T_K = ( \sqrt{2 U \Gamma}/\pi ) \,
      \exp \! \left [ \pi
                 \left (
                        \Gamma^2 + \epsilon_0\,
                         U + \epsilon^2_0
                 \right )
                 /(2\,U\,\Gamma )
           \right ] \, ,
\label{eq:TKAndersonAccurate}
\end{equation}
where $\Gamma =  \pi \rho |V|^2$. This expression for $T_K$
exactly reproduces Eq.~(6.22) of Ref.~\onlinecite{WiegmannC83}
for the symmetric Anderson model, $\epsilon_0 = -U/2$. It also
coincides with Eq.~(7.11) of Ref.~\onlinecite{WiegmannC83} for
the Kondo temperature of the asymmetric model when $U \gg
\Gamma$. Note that the $\Gamma^2$ term in the exponent is
usually omitted from Eq.~(\ref{eq:TKAndersonAccurate}) on the
basis of it being small. It does in general improve the
estimate for $T_K$.

In the local-moment regime, where
Eq.~(\ref{eq:TKAndersonAccurate}) is valid, the impurity
magnetization of the isotropic Anderson model is dominated by
the universal magnetization curve of
Eq.~(\ref{eq:MKfullWiegmann}) up to fields of the order of $h
\sim \sqrt{\Gamma U} \gg T_K$ (see, e.g., lower left inset to
Fig.~\ref{fig:MethodCompare}). At yet larger fields, $h \gg
\sqrt{\Gamma U}$, the magnetization of the Anderson model can
no longer be described by that of the Kondo model, as charge
fluctuations become exceedingly more important than spin flips.
Rather, $M$ is well described by perturbation theory in
$\Gamma$. Importantly, the asymptotic expansion of
Eq.~(\ref{eq:MKassympt}) properly matches (to leading order in
$\Gamma/U$) the perturbative result~\cite{Gefen04} for $M$ when
$h \sim \sqrt{\Gamma U}$. Thus, the two approaches combine to
cover the entire range in $h$ for the Anderson model.

\subsubsection{Bethe \emph{ansatz} equations for the
               occupancy and magnetization}

The Bethe \emph{ansatz} solution of the Anderson model features
four key quantities, which are the distributions of the charge
and spin rapidities, $\tilde{\rho}_{\text{i/h}}(k)$ and
$\tilde{\sigma}_{\text{i/h}}(\lambda)$, respectively, for the
impurity ($i$) and the host ($h$) band. The total impurity
occupancy and magnetization are expressed as integrals over the
distributions of the charge and spin rapidities for the
impurity:
\begin{eqnarray}
M &=& \frac{1}{2}
      \int_{-\infty}^{B} \!\!
             \tilde{\rho}_{\text{i}}(k) \, d k \, ,
\label{eq:M-Bethe} \\
n_d &=& 1 - \int_{-\infty}^{Q} \!\!
                 \tilde{\sigma}_{\text{i}}(\lambda) \,
                 d \lambda \, .
\label{eq:nd-Bethe}
\end{eqnarray}
The upper limits of integration in Eqs.~(\ref{eq:M-Bethe}) and
(\ref{eq:nd-Bethe}) are determined through implicit conditions
on the corresponding distribution functions for the host band,
\begin{eqnarray}
\frac{h}{2\pi} &=& \int_{-\infty}^{B} \!\!
        \tilde{\rho}_{\text{h}}(k) \, d k \, ,
\label{eq:fixing-B}\\
\frac{U+2 \epsilon_0}{2 \pi} &=& \int_{-\infty}^{Q} \!\!
        \tilde{\sigma}_{\text{h}}(\lambda)\, d\lambda \, .
\label{eq:fixing-Q}
\end{eqnarray}

As for the distributions of the rapidities for the impurity and
the host, these are determined by the same pair of linear
integral equations, only with different inhomogeneous parts:
\begin{widetext}
\begin{align}
& \tilde{\rho}(k) + \frac{d g(k)}{d k}
      \int_{-\infty}^{B} R[ g(k)-g(k') ] \,
                         \tilde{\rho} (k') \, dk'
      + \frac{d g(k)}{d k} \int_{-\infty}^Q
              S[g(k)-\lambda ] \, \tilde{\sigma}(\lambda)
              d \lambda = \tilde{\rho}^{(0)}(k) \, ,
\label{eq:rhoBethe} \\
& \tilde{\sigma}(\lambda) - \int_{-\infty}^{Q}
      R(\lambda-\lambda') \, \tilde{\sigma}(\lambda') \,
      d \lambda'
      + \int_{-\infty}^{B} S[\lambda-g(k)] \,
              \tilde{\rho}(k) \, d k =
      \tilde{\sigma}^{(0)}(\lambda) \, ,
\label{eq:sigmaBethe}
\end{align}
\end{widetext}
where~\cite{Comment-on-Rx}
\begin{align}
S(x) &= \frac{1}{2 \cosh(\pi x)} \, ,
\\
R(x) &= \frac{1}{2 \pi} \re \!
         \left [
                 \Psi \left ( 1 + i\frac{x}{2} \right)
                 \!-\! \Psi \left ( \frac{1}{2} +
                                  i\frac{x}{2} \right)
         \right ] ,
\\
g(k) &= \frac{(k-\epsilon_0-U/2)^2}{2 U \Gamma}
\end{align}
(here $\Psi$ is the digamma function). The inhomogeneous parts
in Eqs.~\eqref{eq:rhoBethe} and \eqref{eq:sigmaBethe} are given
in turn by
\begin{align}
\tilde{\rho}^{(0)}_{\text{i}}(k) &=
      \tilde{\Delta}(k) + \frac{d g(k)}{d k}
            \int_{-\infty}^{+\infty} \!\!\! R[ g(k)-g(k') ]
                 \, \tilde{\Delta} (k') \, dk' \, ,
\\
\tilde{\rho}^{(0)}_{\text{h}}(k) &= \frac{1}{2\pi}
      \left \{
               1 + \frac{d g(k)}{d k}
               \int_{-\infty}^{+\infty} \!\!\!
                    R[g(k)-g(k')] \, dk' \right \} \, ,
\end{align}
\begin{align}
\tilde{\sigma}^{(0)}_{\text{i}}(\lambda) &=
      \int_{-\infty}^{+\infty} \!\!\! S[\lambda-g(k)]
               \, \tilde{\Delta}(k)  \,  d k \, ,
\\
\tilde{\sigma}^{(0)}_{\text{h}}(\lambda) &=
      \frac{1}{2\pi} \int_{-\infty}^{+\infty} \!\!\!
               S[\lambda-g(k)] \, d k \, ,
\end{align}
where $\tilde{\Delta} (k)$ is the Lorentzian function
\begin{equation}
\tilde{\Delta}(k) = \frac{1}{\pi}
      \frac{\Gamma}{\Gamma^2 + (k-\epsilon_0)^2} \, .
\end{equation}

\subsubsection{Details of the numerical procedure}

The main obstacle faced with in a numerical solution of the
Bethe \emph{ansatz} equations is the self-consistent
determination of the upper integration bounds that appear in
Eqs.~(\ref{eq:M-Bethe})--(\ref{eq:sigmaBethe}). These are
computed iteratively according to the scheme
\begin{equation}
\tilde{\rho}^{(n-1)}_{\text{h}} \, ,
\tilde{\sigma}^{(n-1)}_{\text{h}} \,
\Rightarrow B^{(n)}\, , Q^{(n)} \,
\Rightarrow \tilde{\rho}^{(n)}_{\text{h}} \, ,
\tilde{\sigma}^{(n)}_{\text{h}} \, .
\end{equation}
Starting with $\tilde{\rho}^{(n-1)}_{\text{h}}$ and
$\tilde{\sigma}^{(n-1)}_{\text{h}}$ as input for the $n$th
iteration, $B^{(n)}$ and $Q^{(n)}$ are extracted from
Eqs.~\eqref{eq:fixing-B} and \eqref{eq:fixing-Q}. Using the
updated values for $B$ and $Q$,
$\tilde{\rho}_{\text{h}}^{(n)}(k)$ and
$\tilde{\sigma}_{\text{h}}^{(n)}(\lambda)$ are then obtained
from the solution of Eqs.~\eqref{eq:rhoBethe} and
\eqref{eq:sigmaBethe}. This cycle is repeated until convergence
is reached. The first iteration in this procedure is usually
initialized with $\tilde{\rho}^{(0)}_{\text{h}}(k)$ and
$\tilde{\sigma}^{(0)}_{\text{h}}(\lambda)$ as input. Standard
techniques are then used to ensure rapid convergence of the
iterative solution. Typically $15$ to $30$ iterations are
required to achieve a relative accuracy of $10^{-4}$ for the
vector $(B,Q)$.

The core of this cycle is the solution of
Eqs.~\eqref{eq:rhoBethe} and \eqref{eq:sigmaBethe}. These are
solved (for given values of $B$ and $Q$) by discretizing the
integration interval with adaptively chosen $500 \div 1000$
mesh points. Once a self-consistent solution is reached for
$B$, $Q$, $\tilde{\rho}_{\text{h}}(k)$ and
$\tilde{\sigma}_{\text{h}}(\lambda)$, the corresponding
distributions of rapidities for the impurity are obtained from
a single solution of Eqs.~\eqref{eq:rhoBethe} and
\eqref{eq:sigmaBethe}. The impurity occupancy and magnetization
are calculated in turn from Eqs.~\eqref{eq:M-Bethe} and
\eqref{eq:nd-Bethe}.

To test the accuracy of our numerical results, we have
extensively checked them against the analytical solution for
the zero-field occupancy $n_d(h=0)$ and the zero-field
susceptibility $dM/dh|_{h = 0}$. In suitable parameter regimes,
we have also compared our results to perturbation theory in
both $U$ and $\Gamma$. In all cases tested the relative errors
in $n_d$ and $M$ were less than $0.05\%$ and $0.5\%$,
respectively. This accuracy can be systematically improved by
increasing the number of discretization points used in solving
Eqs.~\eqref{eq:rhoBethe} and \eqref{eq:sigmaBethe} for the
distributions. Our results were also in full agreement with
those reported by Okiji and Kawasaki,~\cite{Okiji82} except for
$M(h)$ where up to $10\%$ differences were found. Considering
the extensive set of checks that were applied to our results,
it appears that the discrepancy is due to lower numerical
accuracy in the solution of Ref.~~\onlinecite{Okiji82}.


\begin{thebibliography}{54}
\bibitem[{Com({\natexlab{a}})}]{Comm-on-equal-rho}
\bibinfo{note}{Assigning the same density of states, $\rho$, to both leads does
  not restrict the generality of our model, since the case of $\rho_L \neq
  \rho_R$ can be accounted for by an appropriate rescaling of the tunnelling
  matrix $\hat A$.}
\bibitem[{Com({\natexlab{b}})}]{Comment-on-FM}
\bibinfo{note}{For ferromagnetic leads, the conduction-electron density of
  states, $\rho_{\sigma}$, is spin dependent, which can be converted into
  spin-dependent tunnelling amplitudes $V_{\sigma}$ by appropriate rescaling.}
\bibitem[{Com({\natexlab{c}})}]{Comment-on-uniqueness}
\bibinfo{note}{If either $\det \hat{A} = 0$, $h \sin\theta = 0$, or
  $V_{\uparrow} = V_{\downarrow}$, then there is an additional degree of
  freedom in constructing the unitary matrices $R_l$ and $R_d$. As detailed in
  Appendix~\ref{App:SVDdetails}, when $\det \hat{A} = 0$ the angle $\chi_A$ can
  take arbitrary values, while for $h \sin \theta = 0$ the angle $\chi_d$ can
  be arbitrary [see Eqs.~(\ref{eq:phased}) and (\ref{eq:phasel}) for the
  definitions of these angles]. If $V_{\uparrow} = V_{\downarrow}$, then the
  angle $\theta$ can take any value in the range $0 \leq \theta \leq \pi$.}
\bibitem[{Com({\natexlab{d}})}]{Comment-on-phi}
\bibinfo{note}{The phase $\varphi$ is ill-defined when any of the coefficients
  $a_{L1}$, $a_{L2}$, $a_{R1}$, and $a_{R2}$ is zero.}
\bibitem[{Com({\natexlab{e}})}]{Comment-on-Langreth}
\bibinfo{note}{Note that the Friedel-Langreth sum rule was derived for the
  Anderson impurity model rather than the Kondo model.}
\bibitem[{Com({\natexlab{f}})}]{Comment-on-M_K}
\bibinfo{note}{The universal function $M_K(h/T_K)$ pertains, strictly speaking,
  to a global magnetic field that couples simultaneously to the impurity and
  the conduction electrons with equal $g$-factors. Here, by contrast, one is
  dealing with a local magnetic field that couples solely to the impurity.
  However, in the scaling regime of interest here, the impurity magnetization
  is the same for a local and a global magnetic field.}
\bibitem[{Com({\natexlab{g}})}]{Comment-on-charge-fluc}
\bibinfo{note}{Formally, these corrections to the dot charge follow
  quantitatively from the fact that, when properly transformed under the
  Schrieffer-Wolff transformation, the local dot operators
  $d^{\dagger}_{\sigma}$ are slightly mixed with the conduction-electron
  degrees of freedom. Note also that to first order in $\Gamma_{\alpha}$ the
  total charge $n_{\text{tot}}$ is independent of basis rotation.}
\bibitem[{com()}]{comm-Sindel-feature}
\bibinfo{note}{Only small nonmonotonicities in the charging are reported in the
  nonperturbative numerical renormalization-group calculations of
  Ref.~\onlinecite{Sindel05} for $\Gamma_{\spinup} = \Gamma_{\spindown}$. The
  regime of nearly degenerate levels, which is required for the appearance of
  the large feature reported here, has not been studied in
  Ref.~\onlinecite{Sindel05}.}
\bibitem[{Med()}]{MedenThanks}
\bibinfo{note}{The improved fRG data for the inset have been kindly provided to
  us by V. Meden.}
\bibitem[{Rel()}]{Relating-the-Gamma's}
\bibinfo{note}{For the special case where $a_{L 1} = a_{R 1}$ and $a_{L 2} =
  -a_{R 2}$, our $\Gamma_{\uparrow}$ and $\Gamma^{}_{\downarrow}$ are related
  to those of Ref.~\onlinecite{Meden06PRL} though $\Gamma^{}_{\uparrow} = 2
  \max \{ \Gamma^{L}_1, \Gamma^{L}_2 \}$ and $\Gamma^{}_{\downarrow} = 2 \min
  \{ \Gamma^{L}_1, \Gamma^{L}_2 \}$.}
\bibitem[{Com({\natexlab{a}})}]{Comment-on-renormalized-levels}
\bibinfo{note}{Here we have neglected exponentially small corrections of order
  $\Gamma_{\text{tot}} \delta\epsilon_{\text{CIR}}/U^2$ to the total dot
  occupancy. These stem from the fact that $\epsilon_0$ is sightly removed from
  the middle of the Coulomb-blockade valley, i.e., $\epsilon_0 \to \pm
  \delta\epsilon_{\text{CIR}}$ in Eq.~(\ref{eq:n0PT}).}
\bibitem[{Com({\natexlab{b}})}]{Comment-on-Rx}
\bibinfo{note}{To evaluate $R(x)$, we have performed analytic integration of
  Eq.~(4.35) of Ref.~\onlinecite{WiegmannC83}.}
\end{thebibliography}

\end{document}